\documentclass[twocolumn,fleqn,english,aps,prd,preprintnumbers,showpacs,showkeys,nofootinbib,superscriptaddress,floatfix,tightenlines]{revtex4-1}
\usepackage[T1]{fontenc}
\usepackage[utf8]{inputenc}
\usepackage{amsmath}
\usepackage{amsthm}
\usepackage{amssymb}
\usepackage{graphicx}
\usepackage{amsfonts}
\usepackage{amscd}
\usepackage{amsxtra}
\usepackage{epstopdf}
\usepackage{color}
\usepackage{dcolumn}
\usepackage{bm}
\usepackage{multirow}
\usepackage{slashed}

\makeatletter

\usepackage[normalem]{ulem}
\usepackage[dvipsnames]{xcolor}
\usepackage{array}
\usepackage{slashed}
\renewcommand{\sout}{\bgroup \color{red} \ULdepth=-.5ex \ULset}

\makeatother

\normalsize

\usepackage{babel}
\begin{document}
\preprint{INHA-NTG-02/2021}
\title{Baryonic matter and the medium modification of the baryon masses} 
\author{Nam-Yong Ghim}
\email{Namyong.ghim@inha.ac.kr}
\affiliation{Department of Physics, Inha University, Incheon 22212,
   Korea} 

\author{Ghil-Seok Yang}
\email{ghsyang@ssu.ac.kr}
\affiliation{Department of Physics, Soongsil University, Seoul 06978,
   Korea}

\author{Hyun-Chul Kim}
\email{hchkim@inha.ac.kr}
\affiliation{Department of Physics, Inha University, Incheon 22212,
  Korea} 
\affiliation{School of Physics, Korea Institute for Advanced Study 
  (KIAS), Seoul 02455,  Korea}

\author{Ulugbek Yakhshiev}
\email{yakhshiev@inha.ac.kr}
\affiliation{Department of Physics, Inha University, Incheon 22212,
  Korea}
\affiliation{Theoretical Physics Department, National University
of Uzbekistan, Tashkent 700174,
  Uzbekistan}

\selectlanguage{english}%
\begin{abstract}
We investigate the properties of baryonic
matter within the framework of the in-medium modified chiral soliton 
model by taking into account the effects of surrounding baryonic  
environment on the properties of in-medium baryons. 
The internal parameters of the model are determined based on nuclear 
phenomenology at nonstrange sector and fitted by reproducing 
nuclear matter properties near the saturation point. We discuss the 
equations of state in different nuclear environments such as
symmetric nuclear matter, neutron and strange matters.  We 
show that the results for the equations of state are in good agreement
with the phenomenology of nuclear matter.  We also discuss how the
SU(3) baryons masses undergo changes in these various types of nuclear
matter.  
\end{abstract}
\maketitle

\section{Introduction}
It is of paramount importance to understand how the masses of hadrons
undergo changes in nuclear medium, since it is deeply rooted in the
restoration of chiral symmetry and even 
the quark confinement in quantum chromodynamics
(QCD)~\cite{Drukarev:1991fs, Birse:1994cz, Brown:1995qt,
  Saito:2005rv}. As discussed in Ref.~\cite{Drukarev:1991fs}, the
chiral condensate is known to be 
modified in nuclear matter, which reveals the mechanism as to how 
the spontaneous broken chiral symmetry is restored as the nuclear
density increases. This also implies the changes of hadron
masses in it, since the dynamical quark mass arises as a consequence 
of the spontaneous breakdown of chiral symmetry.  
Thus, understanding the medium modification of the nucleon mass has
been one of the most significant issues well over
decades~\cite{Serot:1984ey}. Experimental data also indicate that
the nucleon is modified in nuclei~\cite{Aubert:1983xm,
  Strauch:2002wu, Agakishiev:2014moo, 
  Malace:2014uea, Eskola:2016oht, Kolar:2020mvb}. This means that
other baryons may also undergo changes in nuclear
medium~\cite{Osterfeld:1991ii, Lenske:2018bgr, Knorren:1995ds,
  Papazoglou:1997uw,Wang:2001hw, Lattimer:2012nd, Vidana:2018bdi}.  
When one considers the medium modification of baryons, one should keep
in mind that nuclear matter itself is also affected self-consistently
by the changes of baryons. However, it is very difficult to
relate the medium modification of baryons to nuclear medium consistently,
even in the isolated case of normal nuclear matter. 
 
In the present work, we investigate the medium modification of the
low-lying SU(3) baryons in symmetric matter,  asymmetric matter,
neutron matter, and strange baryonic matter consistently, based on a
pion mean-field approach~\cite{Diakonov:1987ty}. The general idea is
based on the seminal paper by E. \,Witten~\cite{Witten:1979kh,
  Witten:1983}.  In the large $N_c$ (the number of colors) limit, the
nucleon can be viewed as a state of $N_c$ valence quarks bound by the
meson mean fields that is produced self-consistently by the presence
of the $N_c$ valence quarks, since the mesonic quantum fluctuations
are suppressed by the $1/N_c$ factor. This approach has been
successfully applied for describing the various properties of both light
and singly heavy baryons in a unifying manner~\cite{Yang:2010id,
  Yang:2010fm, Yang:2015era, Yang:2016qdz, Kim:2017jpx,Kim:2017khv,
  Yang:2018idi, Yang:2018uoj, Yang:2019tst, Yang:2020klp}. The main
idea of the pion mean-field approach is not to compute dynamical
parameters within the chiral quark-soliton
model~\cite{Diakonov:1987ty, Christov:1995vm}, which realizes 
the pion mean-field approach explicitly, but to fix all relevant
dynamical parameters by using the experimental data. For example, the
masses of the baryon decuplet can be predicted by using the
experimental data on those of the baryon octet and the mass of the
$\Omega$ baryon~\cite{Yang:2010fm}. Actually, this method was already
used in the Skyrme model long time ago~\cite{Adkins:1984cf}. 

The pion mean-field approach can be also extended to the description
of light and singly heavy baryons in nuclear medium. However, since
the model is based on the quark degrees of freedom, one should
consider the quark chemical potential~\cite{Berger:1996hc}, which
means that it is rather difficult to connect the results from this
approach directly to the properties of the baryons in nuclear
matter. Thus, we will follow a variational approach that was adopted
in the medium modified Skyrme 
models~\cite{Rakhimov:1996vq, Yakhshiev:2010kf}. In
these modified Skyrme models various properties of the nucleon and
$\Delta$ isobar have been described in nuclear
matter~\cite{Yakhshiev:2010kf, Kim:2012ts,Jung:2013bya,Hong:2018sqa},
and in finite
nuclei~\cite{Yakhshiev:2001ht,Yakhshiev:2004ui,Meissner:2008mr}. 
The model enables one also to investigate nuclear matter
properties~\cite{Jung:2012sy,Yakhshiev:2013eya,Yakhshiev:2015noa}. 

Thus, we will show in this work how the pion mean-field approach 
can be extended to the investigation of the SU(3) baryon properties in
both nuclear and strange baryonic environments. This can be achieved by  
introducing the density-dependent functionals as variational
parameters. The density functionals will be parametrized and fitted  
completely in the SU(2) sector by taking into account available
experimental and empirical data, the linear-response approximation
being emphasized. This enables us to describe the
strange baryonic matter and properties of baryons in different media
(isospin symmetric, asymmetric and strange baryonic matter).

The present paper is organized as follows. In
Section~\ref{sec:2}, we briefly review the pion mean-field approach,   
discussing the collective Hamiltonian and SU(3) baryon states in free 
space. Then we will proceed to consider a possible modification 
of the model in order to take into account the influence of the surrounding 
baryon environment on the properties of a single baryon in medium. In
Section~\ref{sec:3}, we discuss the results for the binding energy 
in symmetric matter and determine the variational parameters. 
The discussion of the properties of baryons in nuclear and strange
nuclear matter will be followed in Section~\ref{sec:4}. We will also
show how to fit the remaining part of the 
parameters. Then, we are able to discuss the properties of an arbitrary
baryonic matter and the numerical results for the medium modifications
of SU(3) baryons.
The final Section\,\ref{sec:5} is devoted to the summary and conclusion of the
present work and will give an outlook for future investigations.
Some details of the model are compiled in Appendix~\ref{app:A}.

\section{General formalism}
\label{sec:2}

In the pion mean-field approach, the dynamics of the valence and sea
quarks generates the chiral-quark soliton with hedgehog
symmetry~\cite{Pauli:1942kwa, Skyrme:1962vh, Witten:1983,
  Diakonov:1997sj}. Hedgehog symmetry can be regarded as the
minimal generalization of spherical symmetry, which can keep the pion
mean fields effectively~\cite{Diakonov:1997sj}. We are able to derive
the effective collective Hamiltonian by considering the zero-mode
quantization with hedgehog symmetry, taking into account the
rotational $1/N_c$ corrections and the strange current-quark
corrections from the explicit breaking of flavor SU(3) symmetry. 
Note that in the present approach the presence of $N_c$ valence
quarks constrains the right hypercharge $Y'=N_c/3$, which picks up
safely the lowest allowed representations such as the baryon octet
($\bm{8}$) and decuplet ($\bm{10}$).  On the other hand, $Y'$ is
constrained by the Wess-Zumino-Witten term in the SU(3) Skyrme
model~\cite{Guadagnini:1983uv, Mazur:1984yf, Jain:1984gp}. 
In this section, we will directly start from the collective
Hamiltonian. For a detailed derivation, we refer to
Ref.~\cite{Blotz:1992pw} (see also a review~\cite{Christov:1995vm}). 
\subsection{Collective Hamiltonian and SU(3) baryon state}
If we consider both the explicit breakdowns of flavor SU(3)
symmetry and isospin symmetry, we have four different contributions to
the collective Hamiltonian, given as follows:  
\label{subsec:CH}
\begin{align}
H  = M_{{\rm cl}}
\;+\;H_{\mathrm{rot}}
\;+\;H_{\mathrm{sb}} + \; H_{\mathrm{em}},
\label{eq:collH}
\end{align}
where $M_{\mathrm{cl}}$, $H_{\mathrm{rot}}$, and $H_{\mathrm{sb}}$ 
denote respectively the classical soliton mass, the $1/N_{c}$
rotational and symmetry-breaking corrections including the effects of
isospin and flavor $\mathrm{SU(3)_{f}}$ symmetry
breakings~\cite{Blotz:1994pc, Yang:2010fm}. The last term
$H_{\mathrm{em}}$ stands for the term arising from the 
isospin symmetry breaking caused by the electromagnetic
self-energies~\cite{Yang:2010id}. We can
neglect the modification of the electromagnetic self-energies in
nuclear matter~\cite{Meissner:2008mr}. The classical energy arises
from the $N_c$ valence quarks in 
the pion mean fields and the sea quarks coming from the vacuum  
polarization in the presence of the $N_c$ valence quarks:
$E_{\mathrm{cl}}=N_c E_{\mathrm{val}} + E_{\mathrm{sea}}$. By
minimizing $E_{\mathrm{cl}}$ with respect to the
pion fields, we get the pion mean-field solution
self-consistently, which yields the classical soliton mass
$M_{\mathrm{cl}}$.  

The rotational $1/N_c$ corrections, i.e., $H_{\mathrm{rot}}$, can
be derived by the zero-mode collective quantization, since the zero
modes are not at all small, one should take into account them
completely. Regarding the angular velocities of the chiral soliton as
small parameters, we can expand the quark propagator perturbatively in
terms of the angular velocities, we find the rotational $1/N_c$ term 
$H_{\mathrm{rot}}$ as 
\begin{align}
H_{\mathrm{rot}} 
& =  
\frac{1}{2I_{1}}\sum_{i=1}^{3}\hat{J}_{i}^{2}
\;+\;\frac{1}{2I_{2}}\sum_{p=4}^{7}\hat{J}_{p}^{2}.
\label{eq:rotH}
\end{align}
This Hamiltonian depends on two moments of inertia $I_{1,2}$ and expressed in 
terms of the operators $\hat J_i$  corresponding to the generators of
the SU(3) group.  $I_{1}$ and $I_{2}$ give the splitting between
different representations of the SU(3) group. The symmetry breaking part
of the Hamiltonian has the following form 
\begin{align}
H_{\mathrm{sb}} & =  
\left(m_{\mathrm{d}}-m_{\mathrm{u}}\right)
\left(\frac{\sqrt{3}}{2}\,\alpha\,D_{38}^{(8)}(\mathcal{A})
\;\right.\cr
&\left.+\;\beta\,\hat{T_{3}}\;+\;\frac{1}{2}
\,\gamma\sum_{i=1}^{3}D_{3i}^{(8)}(\mathcal{A})
\,\hat{J}_{i}\right)
\cr
 &   
+\;\left(m_{\mathrm{s}}-\bar{m}\right)
\left(\alpha\,D_{88}^{(8)}(\mathcal{A})
\;+\;\beta\,\hat{Y}\right.\cr
&\left.+\;\frac{1}{\sqrt{3}}
\,\gamma\sum_{i=1}^{3}D_{8i}^{(8)}(\mathcal{A})
\,\hat{J}_{i}\right),
\label{eq:Hsb}
\end{align}
where $\alpha$, $\beta$, and $\gamma$ depend on the 
moments of inertia that are expressed as
\begin{align}
\alpha
&=-\left(\frac{2}{3}
\frac{\Sigma_{\pi N}}{m_{\mathrm{u}}+m_{\mathrm{d}}}
-\frac{K_{2}}{I_{2}}\right),\cr
\beta&=-\frac{K_{2}}{I_{2}},\;\;\;\;
\gamma=2\left(\frac{K_{1}}{I_{1}}-\frac{K_{2}}{I_{2}}\right).
\label{eq:abg}
\end{align}
Here $K_{1,2}$ represent the anomalous moments of inertia of the
soliton. $m_{\mathrm{u}}$, $m_{\mathrm{d}}$, and
$m_{\mathrm{s}}$ denote the current-quark masses of the up, down, and
strange quarks, respectively. The $\bar{m}$ designates the average
current-quark mass of the up and down quarks. The
$D_{ab}^{(\mathcal{R})}(\mathcal{A})$ indicate the 
SU(3) Wigner $D$ functions in the representation
$\mathcal{R}$. The $\hat{Y}$ and $\hat{T_{3}}$ are the operators of
the hypercharge and the third component of the isospin, respectively.

In the representation $(p,\,q)$ of the SU(3) group, the sum of the
generators can be expressed in terms of $p$ and $q$
\begin{align}
\sum_{i=1}^{8}J_{i}^{2}
=\frac{1}{3}\left[p^{2}\;+\;q^{2}\;+\;p\;q\;+\;3(p+q)\right],
\label{eq:Jsquare}
\end{align}
which yields the eigenvalues of the rotational collective Hamiltonian
$H_{\mathrm{rot}}$ in Eq.~(\ref{eq:rotH}) as follows:
\begin{align}
E_{(p,\,q),\,J} & =  
\frac{1}{2}
\left(\frac{1}{I_{1}}\,-\,\frac{1}{I_{2}}\right)J(J+1)-\frac{3}{8I_{2}}
\cr
&  
+\frac{1}{6I_{2}}\left[p^{2}\,+\,q^{2}\,+\,3(p\,+\,q)+\,p\,q\right]
\label{eq:EpqJ}
\end{align}
A corresponding eigenfunction is called the collective wave
function for a SU(3) baryon with the quantum numbers of flavor 
$\mathcal{F}=(Y,T,T_{3})$ and spin $\mathcal{S}=(Y^{\prime},J,J_{3})$
\begin{align}
\psi_{_{B}}^{\left(\mathcal{R}\right)}\left(\mathcal{A}\right) 
& =  
\sqrt{\mathrm{dim}\left(\mathcal{R}\right)}
\left(-1\right)^{J_{3}+Y^{\prime}/2}
 D_{\mathcal{FS}}^{\left(\mathcal{R}\right)\ast}(\mathcal{A}),
\label{eq:psi}
\end{align}
where $D_{\mathcal{FS}}^{\left(\mathcal{R}\right)\ast}$ are again the
Wigner $D$ functions in a representation $\mathcal{R}$ and
$\mathrm{dim}\left(\mathcal{R}\right)$ designates the corresponding
dimension of the representation $\mathcal{R}$.

Knowing the eigenvalues and eigenfunctions of the SU(3) baryon states,
we can get their masses of which the explicit forms 
are presented in Appendix~\ref{app:A}.
For a detailed formalism relating to the collective Hamiltonian 
and baryon states, we refer the reader to Ref.~\cite{Yang:2010fm},
where all dynamical parameters such as $I_1$, $I_2$, $K_1$, $K_2$,
$\alpha$, $\beta$, and $\gamma$ are determined by using the
experimental data in a ``\emph{model-independent way}'', so that we
can avoid a specific dynamics of the chiral soliton models.
We now turn to how the model can be extended to nuclear medium.
\subsection{Solitons in nuclear matter}
\label{subsec:NM}
Since we have determined all the dynamical parameters by incorporating
experimental information, we will follow the same strategy also in
nuclear matter. We will fix the density-dependent variational
parameters by using the experimental and empirical data on the
properties of nuclear matter.  So, we start from the average energy
$E^{\ast}$ per baryon in a baryonic system\footnote{Asterisks ``$\ast$'' in
  the superscripts denote in-medium modified quantities.}
\begin{align}
\frac{E^{\ast}}{A} & =  
\frac{ZM_{p}^{\ast}+NM_{n}^{\ast}+\sum_{s=1}^{3}N_sM_{s}^{\ast}}{A}\,,
\label{eq:EstarA}
\end{align}
where $Z$ and $N$ are the numbers of protons and neutrons,
respectively, and $N_s$ is the corresponding 
number of the strange baryons with the corresponding strangeness $S$,
$s=|S|$. $A$ stands for the total number of the baryons
$A=Z+N+N_1+N_2+N_3$. 
Having carried out a simple manipulation, we can rewrite $E^*/A$ as
\begin{align}
\frac{E^{\ast}}{A} =  
M_{N}^{\ast}\left(1-\sum_{s=1}^{3}\delta_{s}\right)
+\frac{1}{2}\delta M_{np}^{\ast}
+\sum_{s=1}^{3}\delta_{s}M_{s}^{\ast},
\label{eq:EstarA1}
\end{align}
where $M_{N}^{\ast}=(M_p^{\ast}+M_n^{\ast})/2$ denotes the average mass of
nucleons, $M_{np}^{\ast}\;=\;M_{n}^{\ast}-M_{p}^{\ast}$ designates
the mass differene of the neutron and the proton in medium.
In addition, we introduce the parameter for isospin asymmetry
$\delta=\left(N- Z\right)/A$. $\delta_{s}=N_{s}/A$ represents the
parameter for the strangeness fraction with the corresponding value of  
subscript $s$. We can take $s=1$, $2$ or $3$, depending on the
hyperons with strangeness $S$ we put.   

The binding energy per baryon in a baryonic matter can be defined as the
difference of the medium average energy per baryon $E^*/A$ and the
energy per baryon $E/A$ for the noninteracting baryonic system. If one
takes the number of the baryons to be infinity, which we can call it
the infinite baryon-matter approximation, we express the binding
energy per baryon in terms of the following external parameters: a
normalized baryonic density $\lambda=\rho/\rho_{0}$, the isospin
asymmetry parameter $\delta$, and the strangeness fraction parameter
$\delta_{s}$ with 
given $s$. Consequently, the binding energy is then written as
\begin{align}
\varepsilon(\lambda,\delta,&\,\delta_{1},
  \delta_{2},\delta_{3})   
=  
\frac{E^{\ast}\left(\lambda,\delta,\delta_{1},
      \delta_{2},\delta_{3}\right)-E}{A}\cr  
&=
\Delta M_{N}\left(\lambda,\delta,\delta_{1},
    \delta_{2},\delta_{3}\right)
    \left(1-\sum_{s=1}^{3}\delta_{s}\right) \cr
&+\frac{1}{2}\delta\,\Delta M_{np}
    \left(\lambda,\delta,\delta_{1},
    \delta_{2},\delta_{3}\right) 
    \cr 
&+\sum_{s=1}^{3}\delta_{s}\Delta M_{s}\left(\lambda,
    \delta,\delta_{1},\delta_{2},\delta_{3}\right),
\label{eq:BE}
\end{align}
where $\Delta M_{N}=M_{N}^{\ast}-M_{N}$ denotes the  
isoscalar mass change whereas $\Delta M_{np}=M_{np}^{\ast}-M_{np}$
stands for the neutron-proton mass change in nuclear medium. They are
explicitly expressed in terms of the in-medium modified
functionals of the chiral soliton 
\begin{align}
\Delta M_{N}&= M_{\mathrm{cl}}^{\ast}-M_{\mathrm{cl}}
+E_{(1,1)1/2}^*- E_{(1,1)1/2}\cr
&-D_{1}^*- D_{2}^*+D_{1}+D_{2}\,,\\
\Delta  M_{np} &=d_{1}^*-d_{2}^*-d_{1}+d_{2}\,,
\end{align}
where the explicit expressions for $D_{1,2}$ and $d_{1,2}$ in free
space are given in Appendix~\ref{app:A} (see
Eqs.\,(\ref{eq:d1})-(\ref{eq:cD2})). $D_{1,2}$ represent the linear
$m_s$ corrections of flavor SU(3) symmetry breaking whereas $d_{1,2}$
denote the effects of isospin symmetry breaking. 
They are related to the model 
functionals to be discussed below through $\alpha$,
$\beta$ and $\gamma$ defined in Eq.~(\ref{eq:abg}). 
Note that for the mass differences of the hyperons 
$\Delta M_s=M_s^{\ast}-M_s$ ($s=1,2,3$) we have the different
expressions for the baryon octet and decuplet.  For the moment, let us
concentrate on the strange baryonic medium made of the hyperons in the
baryon octet as in the case of the nonstrange baryons. Thus, we 
adopt the following expressions for $\Delta M_s$ 
\begin{align}
\Delta M_{1}&=\frac{M_{\Lambda}^*+M_{\Sigma}^*}{2}-
\frac{M_{\Lambda}+M_{\Sigma}}{2}\cr
&=M_{\mathrm{cl}}^{\ast}-M_{cl}+
E_{(1,1)1/2}^*-E_{(1,1)1/2}\,,\\
\Delta M_{2}&= M_{\Xi}^*-M_{\Xi}\cr
&=M_{\mathrm{cl}}^{\ast}-
M_{cl}+E_{(1,1)1/2}^*-E_{(1,1)1/2}\cr
&+ D_{2}^*-D_2\,,\\
\Delta M_{3}&= 0\,.
\end{align}

We now discuss how we can modify the dynamical parameters of
the pion mean-field approach in nuclear medium. We follow the strategy
presented in Refs.~\cite{Yakhshiev:2013eya,Yakhshiev:2015noa} and
assume that the dynamical parameters discussed in
Subsection\,\ref{subsec:CH}, {\em i.e.} 
$M_{\mathrm{cl}}$, $I_{1,2}$ and $K_{1,2}/I_{1,2}$, will be modified as follows:
\begin{align}
M_{\mathrm{cl}}&\rightarrow M_{\mathrm{cl}}^{\ast}  = 
M_{\mathrm{cl}}f_{\mathrm{cl}}(\lambda,\delta,\delta_1,\delta_2,\delta_3),
\\
I_{\mathrm{1}}&\rightarrow I_{\mathrm{1}}^{\ast}  =  
I_{\mathrm{1}}f_1(\lambda,\delta,\delta_1,\delta_2,\delta_3),\\
I_{\mathrm{2}}&\rightarrow I_{\mathrm{2}}^{\ast}=
I_{\mathrm{2}}f_2(\lambda,\delta,\delta_1,\delta_2,\delta_3),\\
\left(m_{\mathrm{d}}-m_{\mathrm{u}}\right)&
\frac{K_{1,2}}{I_{1,2}}\rightarrow
E_{\mathrm{iso}}^{\ast} =  
\left(m_{\mathrm{d}}-m_{\mathrm{u}}\right)\cr
&\times\frac{K_{1,2}}{I_{1,2}}f_0(\lambda,\delta,\delta_1,\delta_2,\delta_3),
\label{eq:isosm}
\\
\left(m_{\mathrm{s}}-\bar{m}\right)&
\frac{K_{1,2}}{I_{1,2}}\rightarrow 
E_{\mathrm{str}}^{\ast}  =  
\left(m_{\mathrm{s}}-\bar{m}\right)\cr 
&\times\frac{K_{1,2}}{I_{1,2}}f_{\mathrm{s}}(\lambda,\delta,\delta_1,\delta_2,\delta_3), 
\label{eq:strm}
\end{align}
where $f_{\mathrm{cl}}$, $f_{0,1,2}$, and $f_{\mathrm{s}}$ represent
the functions of nuclear densities for nuclear medium.  
We will parametrize them based on information about nuclear matter
in the next section. One should keep in mind that in general one
can consider density dependencies in a different manner, depending on
the isospin splitting and the mass splittings in different
representations (see Eqs.\,(\ref{eq:isosm}) and\,(\ref{eq:strm})).
As mentioned previously, we assume that the electromagnetic
corrections to the neutron-proton mass difference is only weakly
affected by nuclear medium. So, we will not consider them in the
present work. We ignore also the effects of isospin symmetry breaking  
coming from baryons on the binding energy except for the nucleons,
assuming the small strangeness fraction up to normal nuclear matter
densities. 
\section{Nuclear phenomenology}
\label{sec:3} 
In the present section, we will discuss the results related to
symmetric nuclear matter, isospin asymmetric nuclear matter, and more
general baryonic matter, one by one. We first start with ordinary
symmetric nuclear matter.  
\subsection{Symmetric nuclear matter}
\label{subsec:OSM}
We first consider isospin symmetric and non-strange ordinary
nuclear matter with the external parameters $\delta=0$ and 
$\delta_{s}=0$. Then we can parametrize the density
functions such as $f_{\mathrm{cl}}$, $f_1$ and $f_2$. In consequence,
we are able to determine the values of the corresponding parameters  
phenomenologically. For example, they are related to the properties of
isospin-symmetric nuclear matter near the saturation point, i.e., at
the normal nuclear matter density $\rho_0\sim(0.16-0.17)\,{\rm fm}^{-3}$. We
remind that in the case of isospin-symmetric nuclear matter the
binding energy per unit volume is given by 
\begin{align}
\mathcal{E}  \equiv \varepsilon_{V}\left(\rho\right)\frac{A}{V}
\;=\; \rho_{0}\lambda\,\varepsilon_{V}\left(\lambda\right).
\label{eq:BEsym}
\end{align}
Following the ideas presented in
Refs~\cite{Yakhshiev:2013eya,Yakhshiev:2015noa},  
we choose the parametrization of the three medium functions 
$f_{\mathrm{cl}}$, $f_1$, and $f_2$, which are independent of the
asymmetry parameter $\delta$ and the strangeness fraction parameters
$\delta_s$. Furthermore, we will parametrize them for simplicity in a
linear density-dependent form 
\begin{align}
f_{\mathrm{cl}}(\lambda)=\left(1+C_{\mathrm{cl}}\lambda\right),
  \quad f_{1,2}(\lambda)=\left(1+C_{1,2}\lambda\right).
\label{eq:den_ftn001}  
\end{align}
It is enough to employ this linear-density approximation, since 
the equations of state (EoS) for nuclear matter are well
explained. However, if the density of nuclear matter becomes larger
than the normal nuclear matter density, one may need to consider
higher-order corrections to the parametrization we use. Nevertheless,
we will compute the baryon properties as functions of the nuclear
matter density up to $3\rho_0$ to see how far the linear-density
approximation works well. 

The properties of symmetric nuclear matter near the 
saturation point can be related to the isoscalar nucleon mass  
change in the nuclear medium
\begin{align}
\varepsilon_V(\lambda)= \epsilon(\lambda,0,0,0,0)=
\Delta M_N(\lambda)\,.
\end{align}
This implies that $\alpha$, $\beta$ and $\gamma$ will not be 
changed in symmetric nuclear medium.
We will see that they will come into play when we consider asymmetric
nuclear and strange baryonic matter.
Then we can easily obtain the following formula for the density
dependence of the volume energy 
\begin{align}
\varepsilon_V(\lambda) = 
\mathcal{M}_{\mathrm{cl}}C_{\mathrm{cl}}\lambda
-\frac{3C_1\lambda}{8I_1(1+C_{1}\lambda)}
-\frac{3C_{2}\lambda}{4I_2\left(1+C_{2}\lambda\right)}.
\label{eq:den_ftn002}
\end{align}

We now proceed to calculate the properties of nuclear matter near the
saturation point $\lambda=1$ by expanding the volume energy with
respect to the nuclear density. The expansion coefficients have
clear physical meanings related to the properties of nuclear matter at
the saturation point. They are given as follows: 
\begin{align}
a_V 
& =  \varepsilon_{V}(1),\qquad
P_{0} = 
\rho_{0}\lambda^{2}
\left.\frac{\partial\mathcal{\varepsilon}_{V}(\lambda)}{\partial
\lambda}\right|_{\lambda=1}\,,\cr
K_{0}&  =  
9\lambda^{2}\left.\frac{\partial^{2}\mathcal{\varepsilon}_{V}
(\lambda)}{\partial\lambda^{2}}\right|_{\lambda=1}\,,
\label{eq:EPK}
\end{align}
where $a_V$ denotes the value of the volume energy, $P_0$ stands for
that of the pressure, and $K_0$ represents the compressibility of
nuclear matter at the saturation point. 
The value of the coefficient of the volume term $a_V$ is well known
from the analysis of atomic nuclei according to the semi-empirical
Bethe-Weizs\"{a}ker formula\,\cite{Bethe:1936zz,Weizscker:1935zz}. 
So, we choose the well-known value $a_{V}=-16\;\mathrm{MeV}$.  
The stability of nuclear matter requires the zero value of the
pressure $P_0=0$ at the saturation point. The compressibility of
nuclear matter within various approaches is found to be   
$K_0\sim (290\pm70)$\,MeV~\cite{Sharma:1988zza, Shlomo:1993zz,
  Ma:1997zzb, Vretenar:2003qm, TerHaar:1986ii, Brockmann:1990cn}. 
Based on a comprehensive reanalysis of recent data on the energies of
the giant monopole resonance (GMR) in even-even ${}^{112-124}$Sn and  
${}^{106,100-116}$Cd and earlier data on $58\le A \le 208$ 
nuclei in Ref.\,\cite{Stone:2014wza}, the 
value of the compressibility can be taken to be $K_0\sim (240\pm 
20)$\,MeV~\cite{Shlomo2006}. Following this analysis, 
we choose $K_{0}=240\,\mathrm{MeV}$ in the present work. Thus, 
the three parameters in the density functions given in
Eq.~\eqref{eq:den_ftn001} can be fitted to be  
\begin{align}
C_{\mathrm{cl}} 
=  
-0.0561,
\;\;
C_{1}
=
0.6434,
\;\;
C_{2}
=
-0.1218.
\label{eq:CclC1C2}
\end{align}
Now we can predict the skewness of symmetric nuclear matter, which
is defined from the fourth coefficient in the series of the volume
energy:  
\begin{align}
Q&=27\lambda^{3}\left. \frac{\partial^{3} \varepsilon_{V}
(\lambda)}{\partial \lambda^{3}}\right|_{\lambda=1}\cr
&=-\frac{117}2\left(\frac{C_{1}^{3}}{(1+C_{1})^{4}I_{1}}-\frac{2 
C_{2}^{3}}{(1+C_{2})^{4}I_{2}}\right)\cr
&=-182\ \mathrm{MeV}\,.
\end{align}
The result is consistent with those from other model calculations. For
example, one can find similar results from the Hartree-Fock approach
based on the Skyrme interactions~\cite{Chabanat:1997qh} and the
isospin- and momentum-dependent interaction (MDI) 
model~\cite{Das:2002fr}. We want to emphasize that these coefficients
in the expansion of the volume energy can be used for understanding
the properties of symmetric nuclear matter. 

\subsection{Asymmetric nuclear matter}
\label{subsec:ANM}
Since the asymmetric nuclear matter arises from the isospin symmetry
breaking, Eq.~\eqref{eq:isosm} plays the key role in describing the
asymmetric nuclear matter.  Following the strategy taken from 
Refs.~\cite{Yakhshiev:2013eya, Yakhshiev:2015noa},  
we find that the density function $f_0$ can be defined as a function
of the normalized density $\lambda$ and isospin asymmetry parameter
$\delta$ in the following form 
\begin{align}
f_0(\lambda,\delta)
=1+\frac{C_{\mathrm{num}}\lambda\,\delta}{1+C_{\mathrm{den}}
\lambda}\,,
\label{eq:f0}
\end{align}
where $C_{\mathrm{num}}$ and $C_{\mathrm{den}}$ can be determined
phenomenologically.  This parametrization is also chosen under the
simple assumption: that is, if $\delta$ is zero or
$\rho$ is zero, then the value of $f_0$ is equal to 1.  
Moreover, using the parametrized form given in
Eq.~\eqref{eq:f0}, one can see that various properties of asymmetric
nuclear matter are well described, e.g., the binding energy of
asymmetric nuclear matter will be given as a quadratic form with
respect to the asymmetry parameter $\delta$.\footnote{Note that there 
  is another $\delta$ factor in Eq.\,~(\ref{eq:BE}).}

The nuclear symmetry energy is defined as the second
derivative of the binding energy with respect to $\delta$:
\begin{align}
\varepsilon_{\mathrm{sym}}\left(\lambda\right) 
& =  
\frac{1}{2!}
\left.\frac{\partial^{2}\varepsilon
 \left(\lambda,\delta,0,0,0\right)}{\partial\delta^{2}}
\right|_{\delta=0} .
\label{eq:symen}
\end{align}
As in the case of the volume energy, we can expand
$\varepsilon_{\mathrm{sym}}$ around the saturation 
point  $\lambda=1$ as follows 
\begin{align}
\varepsilon_{\mathrm{sym}}\left(\lambda\right) 
& =  
a_{\mathrm{sym}}+\frac{L_{\mathrm{sym}}}{3}\left(\lambda-1\right)\cr
&+K_{\mathrm{sym}}\frac{\left(\lambda-1\right)^{2}}{18}
+\cdots,
\label{eq:symE}
\end{align}
from which we obtain the value of the nuclear symmetry energy at 
saturation point $a_{\mathrm{sym}}$, that of its slope parameter 
$L_{\mathrm{sym}}$, and the asymmetric part of the compressibility
$K_{\mathrm{sym}}$. They are explicitly written as  
\begin{align}
a_{\mathrm{sym}} 
& =  
-\frac{9}{20}\frac{C_{\mathrm{num}}\left(b-7r/18\right)}
                    {\left(1+C_{\mathrm{den}}\right)},\\ 
L_{\mathrm{sym}}
&=
-\frac{27}{20}\frac{C_{\mathrm{num}}\left(b-7r/18\right)}
                   {\left(1+C_{\mathrm{den}}\right)^{2}},\\ 
K_{\mathrm{sym}}&=\frac{81C_{\mathrm{den}}
                  C_{\mathrm{num}}(b-7r/18)}{10(1+C_{\mathrm{den}})^{3}},  
\label{eq:al}
\end{align}
where  $b= (m_{\mathrm{d}}-m_{\mathrm{u}})\beta$ and
$r=(m_{\mathrm{d}}-m_{\mathrm{u}})\gamma$. 

The value of the nuclear symmetry energy at the saturation point is known
to be in the range $\varepsilon_{\rm sym}(1)\sim 30-34$ MeV. So, we
can take the average value $a_{\mathrm{sym}}=32$\,MeV.  The
correlation between the value of the symmetry energy at the saturation  
density and that of its slop parameter taken from the neutron skip
thickness experiments of  $^{68}$Ni, $^{120}$Sn, and $^{208}$Pb
indicates the tendency that heavier the nucleus yields larger the
value of $L_{\mathrm{sym}} $, which corresponds to that of
$a_{\mathrm{sym}}$~\cite{Roca-Maza:2015eza}. 
As a result, one can choose $L_{\mathrm{sym}}=60$\,MeV for
the asymmetric nuclear matter.  We mainly use these values of 
$a_{\mathrm{sym}}$ and $L_{\mathrm{sym}}$ in the course of the present 
calculation, if it is not specified otherwise.\footnote{We will not
 use any other input data in the strangeness sector.} 
The empirical values of these two quantities will adjust those of 
 $C_{\rm num}$ and $C_{\rm den}$,  respectively.
In order to check the stability of the present results for
to neutron matter ($\delta=1$), however, we have analyzed the
different choices for the $a_{\rm sym}$ and $L_{\rm sym}$ with small
variations. Note that the results are very insensitive to
$a_{\mathrm{sym}}$ in the range of its values discussed above. Thus,
we will only show the variations of $L_{\mathrm{sym}}$ and the
possible two choices of $L_{\mathrm{sym}}$ in this work are listed in
Table~\ref{tab:1}.  
\begin{table}[hbt]
\caption{Possible sets of the parameters for the symmetry energy.}
\label{tab:1}
\begin{tabular}{lcccc}\hline \hline
 & $a_{\rm sym}$ [ MeV] & $L_{\rm sym}$ [MeV] & $C_{\rm num}$ &  $C_{\rm den}$ \\
 \hline
 Set I & 32 & 60 & 65.60 & 0.60\\
 Set II & 32 & 50 & 78.72 & 0.92\\
 \hline\hline
\end{tabular}
\end{table}
All the parameters are actually fitted in this way in relation to
nuclear matter properties at the saturation density, so the present
model can be regarded as a simple model of nuclear matter with five
parameters. Using the values of the parameters for the symmetry energy
listed in Table~\ref{tab:1}, we are able to discuss the EoS for
asymmetric nuclear matter, extrapolating to the low and high density
regions, and to predict various properties of nuclear matter. In
particular, employing Set~I, we can calculate the third coefficient in the
expansion of the symmetry energy, which leads to
$K_{\mathrm{sym}}=-135\ \mathrm{MeV}$. The following quantities, which 
are related to $K_{\mathrm{sym}}$, can be also determined as 
\begin{align}
&K_{\tau}= K_{\mathrm{sym}}-6L_{\mathrm{sym}} = -495\ 
\mathrm{MeV}, \cr
&K_{(0,2)}=K_{\tau}-\frac{Q}{K_{0}}L_{s}=-450\ \mathrm{MeV}.
\end{align}
The calculated values of $K_\tau$ and $K_{0,2}$ are 
in good agreement with the results from other approaches. 
As an example, we can compare the range of $K_{0,2}$ 
value with that from the phenomenological momentum-independent  
model $-477\,\mathrm{MeV} \le K_{0,2}\le
-241\,\mathrm{MeV}$\,\cite{Chen:2009qc}.  
 
\begin{figure}[hbt]
\centering \includegraphics[scale=0.2]{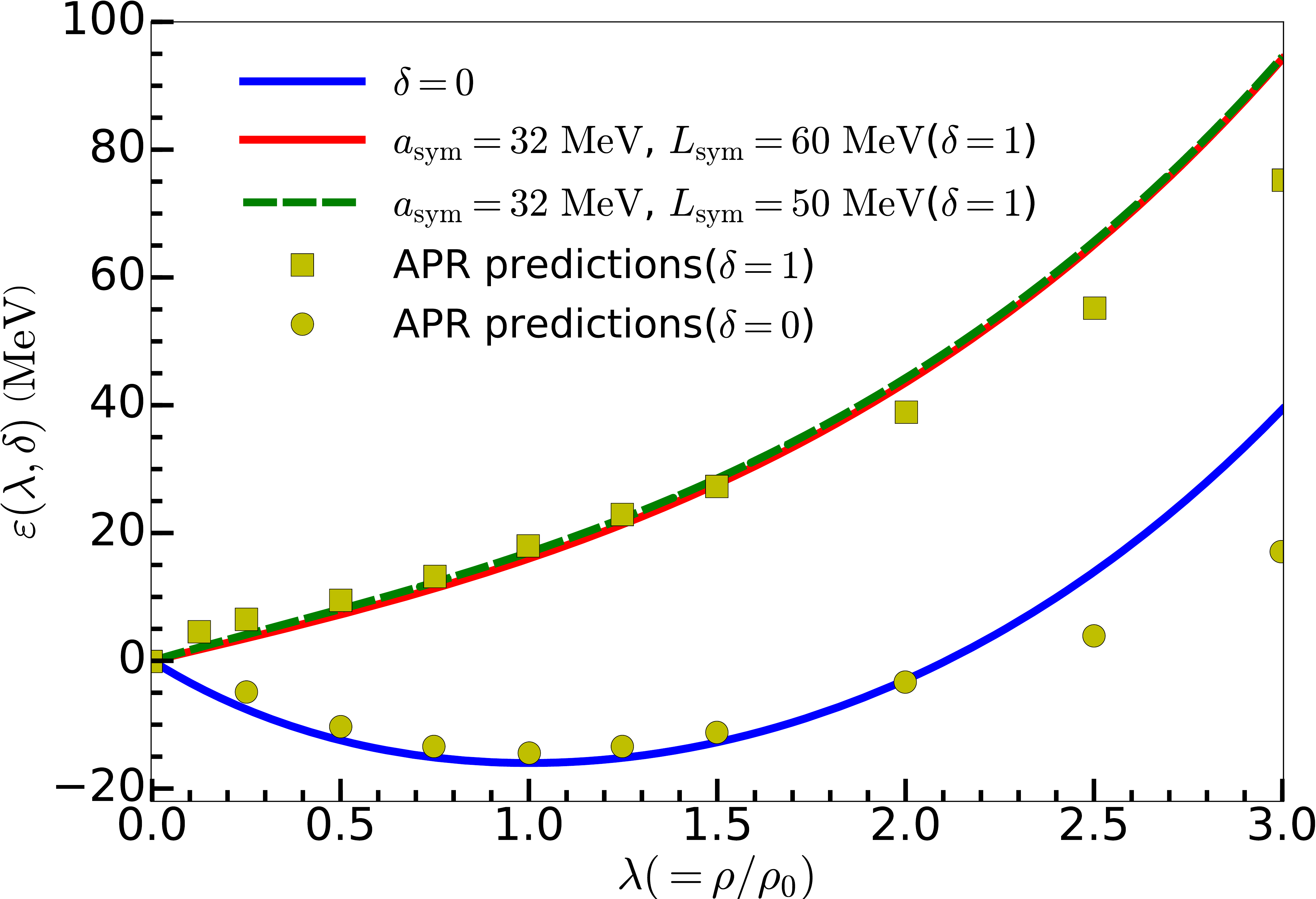}
\caption{Binding energy per nucleon $\varepsilon(\lambda,\delta)\equiv 
\varepsilon(\lambda,\delta,0,0,0)$ as a function of the normalized
 nuclear matter density $\lambda=\rho/\rho_{0}$ in unit of MeV.
The blue solid curve depicts the symmetric matter $\delta=0$ whereas 
the red solid and green dashed curves illustrate those of
neutron matter $\delta=1$ for the possible two sets of
symmetry energy parameters, respectively. 
The present results are compared with those given in APR
predictions~\cite{Akmal:1998cf} that are given by the yellow circles
and boxes, respectively.} 
\label{fig:1} 
\end{figure} 
Figure~\ref{fig:1} draws illustratively the density dependence 
of binding energy per nucleon, given the different values
of the asymmetry parameter $\delta$. We find that the results are
rather stable to the change of  values of the parameters $a_{\rm
  sym}$ and $L_{\rm sym}$. In
particular, the present results change only slightly 
as the values of $L_{\rm sym}$ are varied from 50 to 60\,MeV.

It is natural that the neutron matter gets less bound relatively to
the symmetric matter, as already shown in Fig.~\ref{fig:1}.   
The density dependence of the binding energy per nucleon in symmetric
matter and neutron matter are in agreement with those from other
models and phenomenological ones. In particular, it is consistent with 
Akmal-Pandharipande-Ravenhall (APR) predictions~\cite{Akmal:1998cf} in
the range of $\lambda$, where the simple linear-density approximation
is justified for the medium modification of the corresponding soliton
functionals in nuclear matter. As $\lambda$ increases, the present EoS
becomes stiffer such that one can get the larger masses of 
neutron stars than the solar mass. However, as the density becomes
higher than $\lambda=2$, the linear density approximation may not be 
enough, which requires one to introduce higher-order nonlinear terms. 

\begin{figure}[htp]
\centering \includegraphics[scale=0.2]{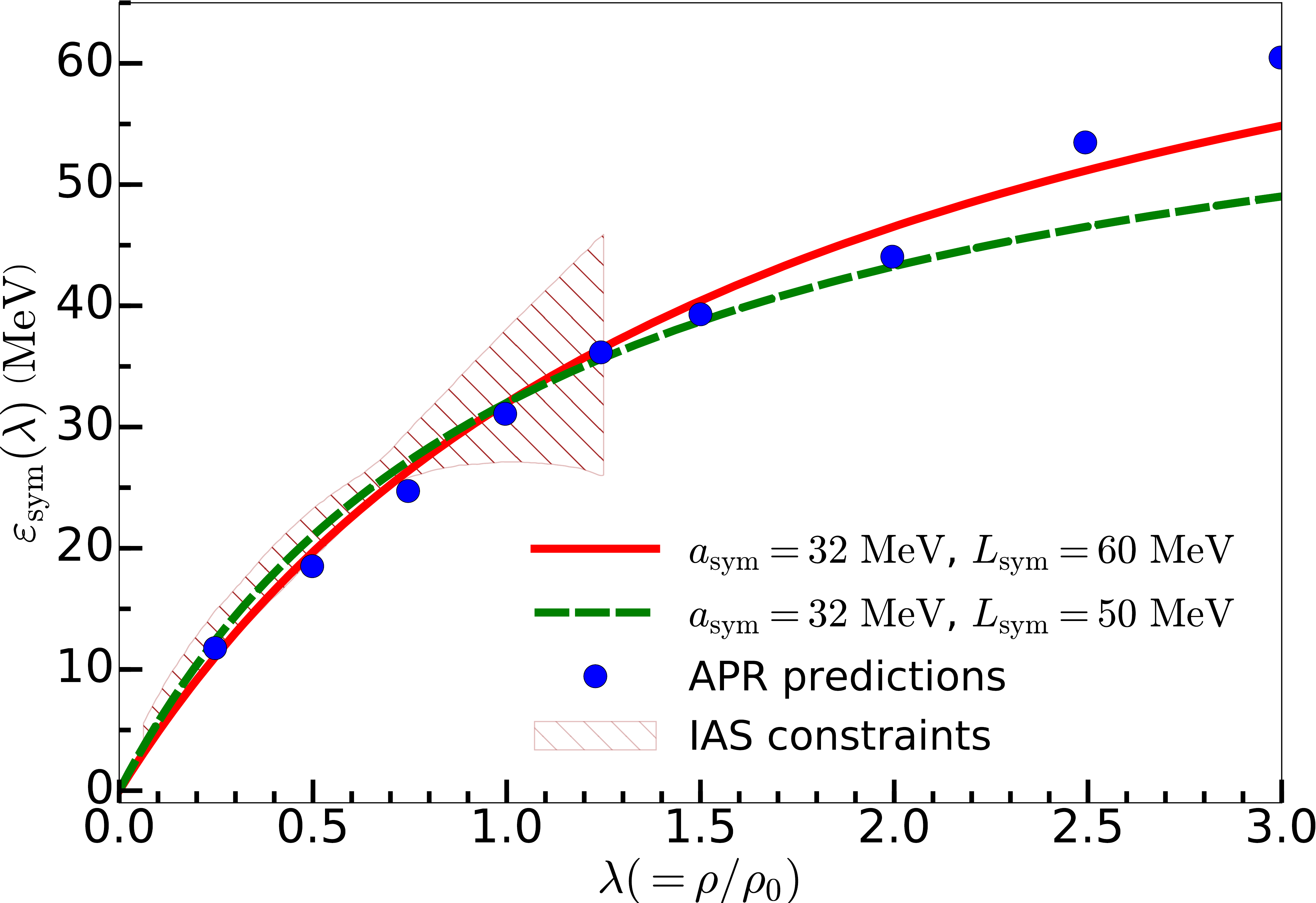}
\caption{Nuclear symmetry energy $\varepsilon_{\mathrm{sym}}
(\lambda)$ as a function of the normalized nuclear density
$\lambda=\rho/\rho_{0}$ in unit of $\mathrm{MeV}$. The results with the
possible two sets of parameters for the symmetry energy are
represented by the 
red solid and green dashed curves,
respectively. The results are compared with those from
Ref.~\citep{Akmal:1998cf}, which are  
marked by the blue circles and those from the IAS
constraints\,\cite{Danielewicz:2013upa} shown 
by the shaded region.} 
\label{fig:2}
\end{figure}
The nuclear symmetry energy plays a very important role in
understanding the EoS of nuclear matter and, in particular, of the
neutron matter. Figure~\ref{fig:2} exhibits how the nuclear symmetry  
energy depends on $\lambda$. We present the results with the two
sets of the parameters $a_{\rm sym}$ and $L_{\rm sym}$ listed in
Table~\ref{tab:1}. When the value of the slop parameter
$L_{\mathrm{sym}}$ gets smaller, the symmetry energy becomes slightly
larger than that obtained by using the larger value of
$L_{\mathrm{sym}}$ till the normal nuclear matter density
($\lambda=1$), then it becomes smaller than that with
$L_{\mathrm{sym}}=60$ MeV. Note that, however, the present results are
quite stable as the parameters vary, and are consistent with those
obtained from other approaches and extracted data.  
In particular, the results are in good agreement with APR predictions
till the density reaches $\lambda =2$. At large nuclear matter 
densities the results of the symmetry energy become smaller than the
values of the APR symmetry energy. The present results are also 
in good agreement with the bounded values of the symmetry energy,
obtained from the analysis of isobaric states 
(IAS)\,\cite{Danielewicz:2013upa}, which is represented  
by the shaded region in Fig.~\ref{fig:2}.

\begin{figure}[htp]
\centering 
\includegraphics[scale=0.2]{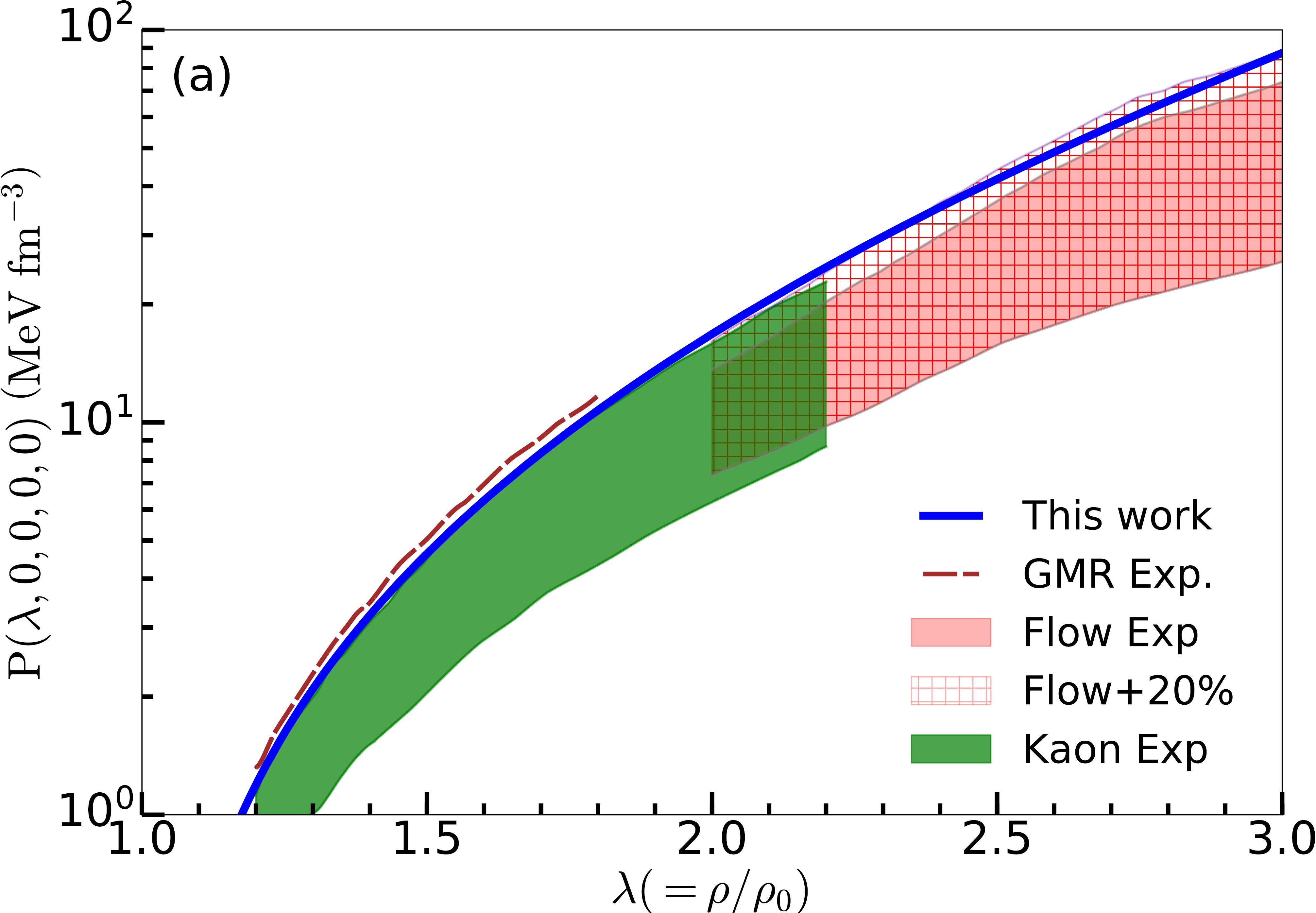}\hspace{0.5cm}
\includegraphics[scale=0.2]{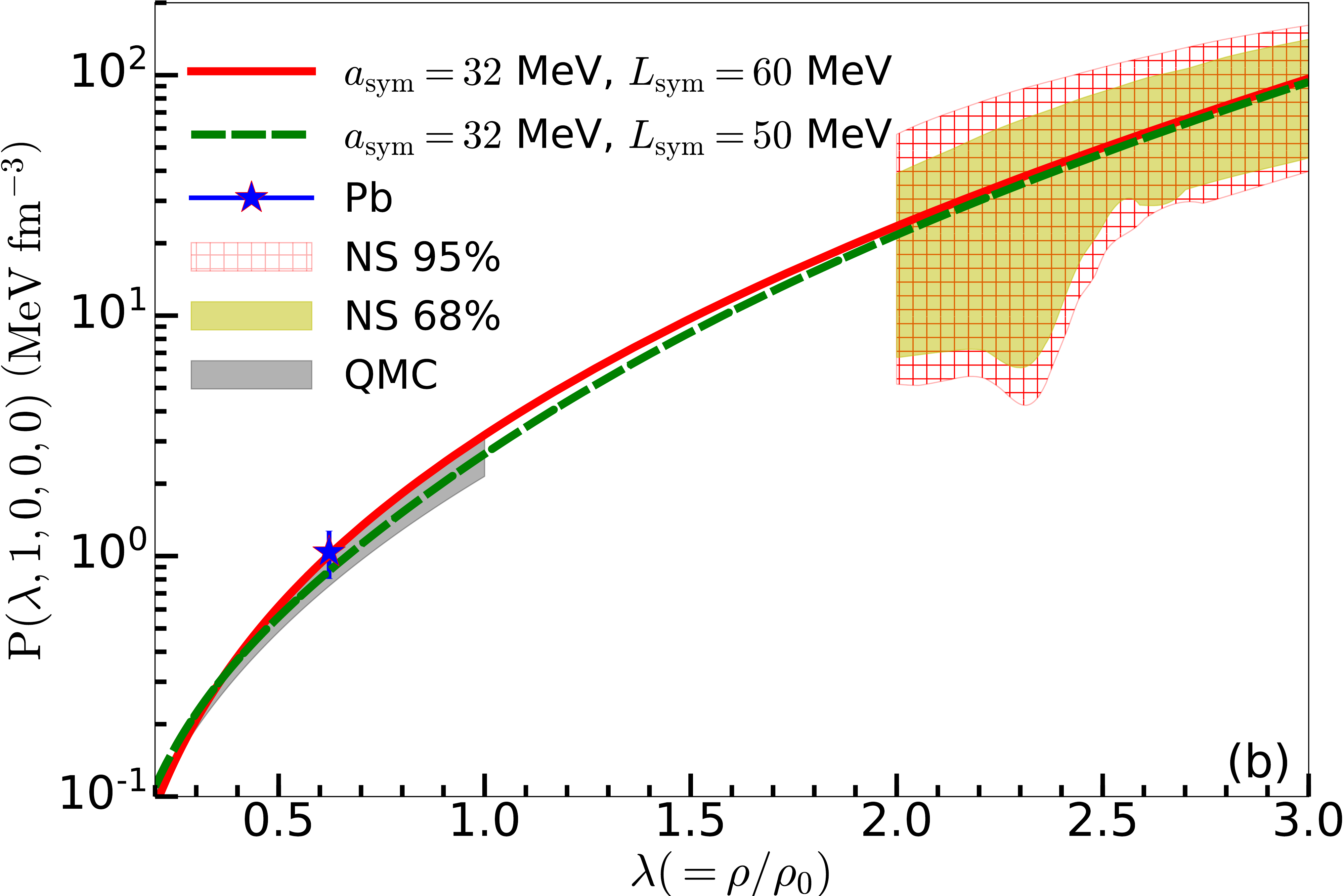}
\caption{Numerical results for the pressure
  $P(\lambda,\,\delta,\,0,\,0,\,0)$. In the upper (a) panel, the
  results are drawn for the symmetric nuclear matter ($\delta = 0$),
  compared with the data taken from
  GMR\,\citep{Youngblood:1999zza,Lynch:2009vc}, 
flow\,\citep{Danielewicz:2002pu}, flow$+20
\%$\,\citep{Steiner:2012xt,Dutra:2014qga}, and
kaon\,\citep{Fuchs:2005zg,Lynch:2009vc} experiments, whereas in the
lower (b) panel those for the neutron 
matter ($\delta = 1$) are depicted, compared with the data from Pb
experiment\,\citep{Tsang:2012se}, NS $95 \%$\,\citep{Steiner:2012xt},
NS $68 \%$\,\citep{Steiner:2012xt}, and Quantum Monte-Carlo
calculations\,\citep{Tsang:2012se,Gandolfi:2011xu,Steiner:2011ft}. In
the neutron matter,  the present results are obtained for the 
two sets of the parameters: red solid and green dashed curves draw the
results with Set I and Set II, respectively.}  
\label{fig:3}
\end{figure}
For completeness, we present in the upper (a) and lower (b) panels of
Fig.~\ref{fig:3} the density dependence of the pressure in the
symmetric nuclear matter and and in the neutron matter, respectively.
The present results for the pressure are in good agreement with those
obtained from other approaches and the extracted data, in
particular, in the range of $\rho\in[0,3\rho_0]$.

For example, in the upper (a) panel of Fig.~\ref{fig:3} the result of
this work for the pressure $P(\lambda,0,0,0,0)$ in the symmetric
matter $(\delta = 0)$, which is drawn in the blue solid curve, is
compared with the data extracted from various 
experiments. In particular, the present result is in good agreement
with the data in the range of $1.2 \leq \lambda \leq 
1.7$ extracted from the GMR
experiments\,\citep{Youngblood:1999zza,Lynch:2009vc}  
for heavy nuclei, which are shown by the dashed curve.
On the other hand, in Ref.\,\citep{Danielewicz:2002pu} 
the flow experimental data on ${}^{197}\mathrm{Au}$ nuclei
collision are analyzed, which are illustrated by the red-shaded region
and correspond to the zero-temperature equation of state for
the symmetric nuclear matter.  Additional studies are presented in
Refs.~\citep{Steiner:2012xt,Dutra:2014qga}, which were extended to the
range of the validity taking into account the mass-radius relation
of neutron stars from observational data.  In the upper panel of
Fig.~\ref{fig:3} data in the extended region are denoted as
"$\mathrm{Flow}+20\%$". One can see that our equations of state are
consistent with the newly predicted range. The EoS for the symmetric
nuclear matter in the range of  $1.2 \leq \lambda \leq 2.2$ can also
be constrained by the kaon production data from high-energy
nucleus-nucleus collision\,\citep{Fuchs:2005zg,Lynch:2009vc}. They are
shown in the green-colored region. The present results lie also within
that region and, in general, are consistent with all the data
extracted from different methods. 

In the lower (b) panel of Fig.~\ref{fig:3}, the results for the
pressure are presented in the neutron matter ($\delta = 1$). We again depict the results for
$P(\lambda,1,0,0,0)$ with the two sets of the parameters $a_{\rm sym}$
and $L_{\rm sym}$. They are represented by the red solid and green dashed
curves, respectively in the lower panel of Fig.~\ref{fig:3}. One can
see that EoS obtained in the present work are quite stable with
the varying parameters that define the symmetric energy. We compare
the results with those extracted from the several experiments. For
example, the weighted average of the experimental data on the neutron
skin thickness in ${}^{208}\mathrm{Pb}$ is indicated by the
red-colored star at subnuclear density\,\citep{Tsang:2012se}. The
studies in Ref.\,\citep{Steiner:2012xt} provide a constraint for the
pressure values of neutron star matter from astrophysical observation
data. In the lower panel of Fig.~\ref{fig:3}, this constraint is
labeled as "$\mathrm{NS}\,95\%$" and "$\mathrm{NS}\,68\%$" for the
two different confidence limits.  There are also results at the
low-density region from the quantum Monte Carlo calculation
(QMC)\,\citep{Tsang:2012se,Gandolfi:2011xu,Steiner:2011ft}.  They are 
denoted by the gray-shaded region. One can see that our results 
are in an excellent agreement with all extracted data in the different
ways in all density regions presented in the figure.  

In short summary, this simple five-parametric model for nuclear matter
within the framework of the model-independent chiral soliton approach
describes the isospin-symmetric and neutron matter properties very
well. This implies that the meson mean-field approach quite
successful not only for explaining various properties of light and
singly-heavy baryons in free space\,\cite{Yang:2010fm} but also for
describing phenomenologically nuclear matter  
properties, based on minimal phenomenological information in the
nonstrange sector.   

\subsection{Baryonic matter}
\label{subsec:BM}
We now proceed to baryonic matter properties in a more general
case taking into account also the strange baryons. So far, we have
concentrated on the nonstrange sector and fitted our parameters
according to the nuclear phenomenology in non-strange sector. We have
also parametrized the influence of surrounding nuclear matter to the
in-medium nucleon properties in such a way that the binding energy per
nucleon appears as a quadratic term in the isospin asymmetry parameter 
$\delta$. Following the strategy used in the nonstrange sector we can
parametrize the influence of baryonic matter with the strangeness
content. In doing that, we will consider the simplicity as a guiding
principle.  Therefore, as a first step we will not introduce any new 
parameter and try to describe the strangeness-mixed baryonic matter.
For that purpose, we expand the binding energy per baryon into the
series in the region with the small values of the isospin-asymmetry   
parameter $\delta$ and strangeness-mixing parameters $\delta_s$.
Consequently, the series of the binding energy per nucleon at the
small values of isospin asymmetry and hyperon mixture parameters
$\delta$ and $\delta_{s}$ ($i=1,2,3$) given in Eq.(\ref{eq:BE}) can be 
written as 
\begin{align}
\varepsilon(\lambda,\delta,&\delta_{1},\dots)  =  
\varepsilon_{V}\left(\lambda\right)
+\varepsilon_{\mathrm{sym}}\left(\lambda\right)\delta^{2}\cr
&+\sum_{s=1}^3
\left.\frac{\partial\,\varepsilon 
\left(\lambda,\,\delta,\,\delta_{1},\dots\right)}{
\partial\,\delta_{s}}\right|_{\delta =\delta_{1}=\dots=0}\delta_{s}\cr 
&+
\frac{1}{2}\sum_{s,p=1}^3\left.\frac{\partial^{2}\, 
\varepsilon\left(\lambda,\,\delta,\,\delta_{1},\dots\right)}{
\partial\,\delta_{s}\partial\delta_{p}}\right|_{\delta=\delta_{1}=\dots=0} 
\delta_{s}\delta_{p}\cr &+\cdots,
\label{eq:Eddi}
\end{align}
where, for convenience of discussion, the  terms of the standard
volume and symmetry energies for ordinary nuclear  
matter are explicitly separated as the first and the 
second ones. It is obvious that the linear terms in $\delta$
are absent due to the quadratic dependence of the binding energy per
baryon on it.  

Next, assuming that the contributions of higher-order terms 
in $\delta_s$ are negligible, we can choose $f_s$'s similar to $f_0$
(see Eq.\,(\ref{eq:f0})) as a linear form in
$\delta_s$.\footnote{Note that this is also the simplest choice.} 
Furthermore, we parametrize $f_s$ in such a way that there is no $\delta$  
dependence. These parametrization will keep all our discussions in
the nonstrange sector intact. Then we have the following forms
of the remaining density functions 
\begin{align}
f_s\left(\lambda,\,\delta,\,\delta_{1},\dots\right) =1+
g_s(\lambda)\delta_s\,.
\end{align}
We also assume that the third term in Eq.\,(\ref{eq:Eddi}) is 
equals to zero. This leads to the following form of $g_s$
\begin{align}
g_s(\lambda) &=  sg(\lambda),\cr
g(\lambda)&=-\frac{5(M_{\mathrm{cl}}^{\ast}-M_{cl}+
E_{(1,1)1/2}^*-E_{(1,1)1/2})}{3(m_s-\hat m)}\cr
&\times\left(6\frac{K_2}{I_2}+\frac{K_1}{I_1}\right)^{-1}\,.
\label{eq:gs}
\end{align}
This final expression is a reasonable one, because the strangeness
content of nuclei is negligible and $g_s$ at small densities  
in Eq.\,(\ref{eq:gs}) maximizes the energy for  $\delta_s=0$.
This choice is advantageous, since it allows one to fit all parameters 
in the SU(2) sector. As a result, we have no internal density
parameters in the SU(3) sector and we do not need to relate this
approach to the strange matter phenomenology.  All results in the
strangeness sector can be considered as predictions in this
simplified work.

In the medium-modified SU(3) sector, we have only one external free 
parameter, which is the fraction of strange matter.  In order not to 
distinguish the species of strange matter, we introduce the
strangeness-mixing parameter $\chi$ defining 
it as the following simple and  reasonable way: $\delta_s=s\chi$.  
So, we can discuss the strangeness effects by considering 
nonzero values of the free parameter $\chi$. 

\begin{figure}[htb]
\centering 
\includegraphics[scale=0.2]{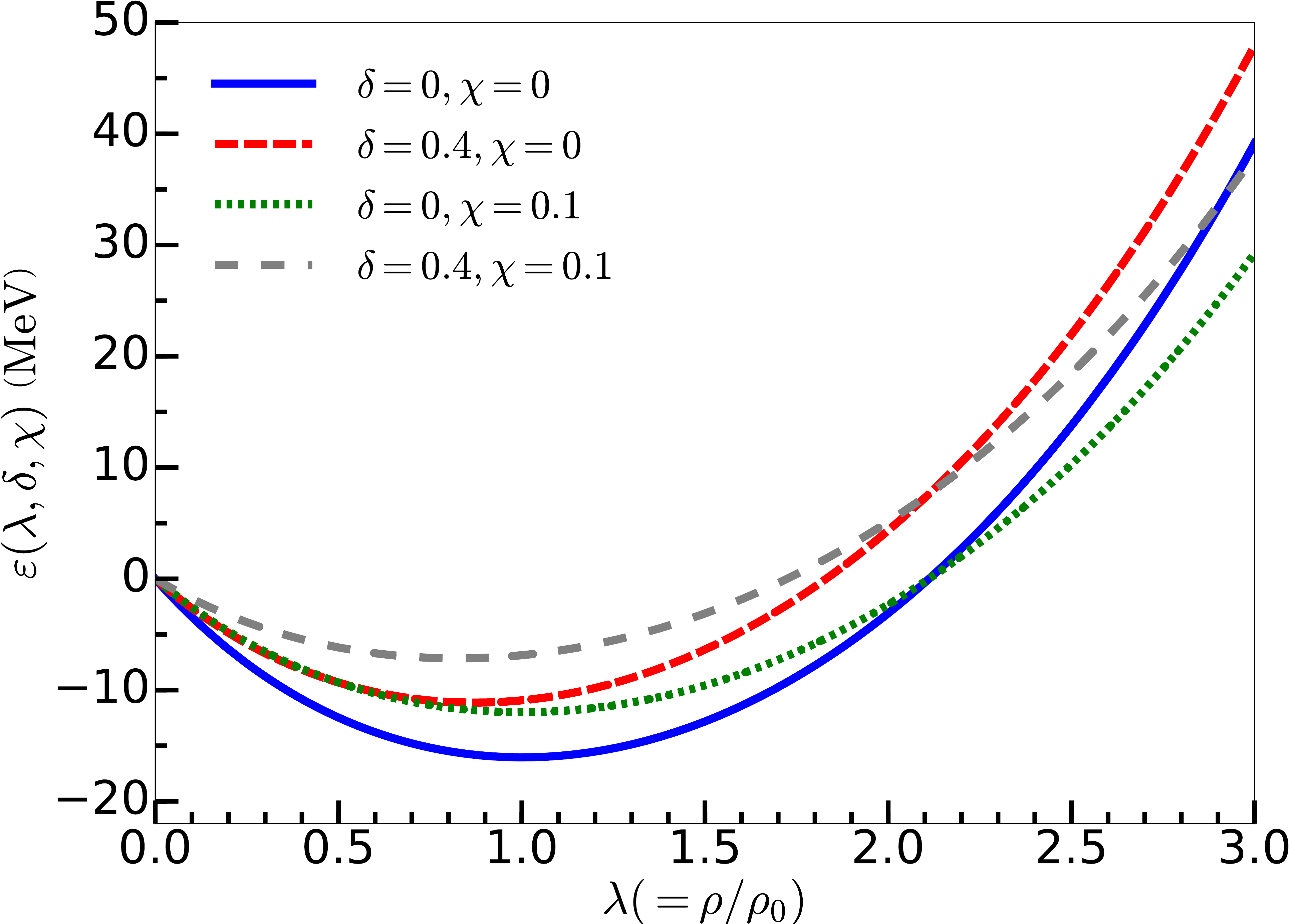}
\caption{Binding energy per nucleon  
$\varepsilon(\lambda,\delta,\chi)$ as a function of the normalized
nuclear matter density $\lambda=\rho/\rho_{0}$.  The results
are drawn for the ordinary isospin symmetric matter in the blue solid
curve, the strangeness mixed isospin-symmetric matter in the red dashed
one, the pure neutron matter in the green dotted one and the
strangeness-mixed isospin-asymmetric matter in the gray space-dashed
one, respectively. The parameters of the symmetry energy are taken
from Set I in Table~\ref{tab:1}.}
\label{fig:4}
\end{figure}
The strangeness effects due to the surrounding environment may come from
the different combinations, e.g. the isospin-symmetric matter with the
strangeness-mixing or the isopin-asymmetric matter with the strangeness
mixing. The binding energies per nucleon for different nuclear
matters are presented in Fig.~\ref{fig:4}, which shows clearly how 
the binding energy undergoes modification as the strangeness content
varies together with $\delta$ changed. For 
comparison, we again depict the binding energy for symmetric
matter in the blue solid curve, for the isospin-asymmetric matter in
the red-dashed one, for the strangeness-mixed isospin-symmetric matter 
in the green-dotted curve and the strangeness-mixed 
isospin-asymmetric matter in the gray space-dashed curve,
respectively. One can see that the strangeness mixing leads to the
less bound system at subnuclear matter densities while 
the binding energy per nucleon varies rather slowly as $\lambda$
increases, so that its magnitude becomes even larger than those in
both iso-symmetric and iso-asymmetric nuclear matter at supranuclear
matter densities, i.e. compare the blue solid 
curve and the green dotted one or the red dashed and grey dashed ones, 
respectively.  At large densities the strange matter may be a more
favorable system so that strange quark stars are allowed to exist with
a smaller mass due to the softening EoS in comparison with neutron
stars. In general, the effect from the isospin asymmetric environment
is much stronger than that from strangeness mixing. The results are
consistent with those from other approaches and model calculations. 
For example, see a recent review~\cite{Lenske:2018bgr} 
about theoretical approaches to the production of 
hyperons, baryon resonance and hyperon matter in 
heavy-ion collision.

\begin{figure}
\centering 
\includegraphics[scale=0.2]{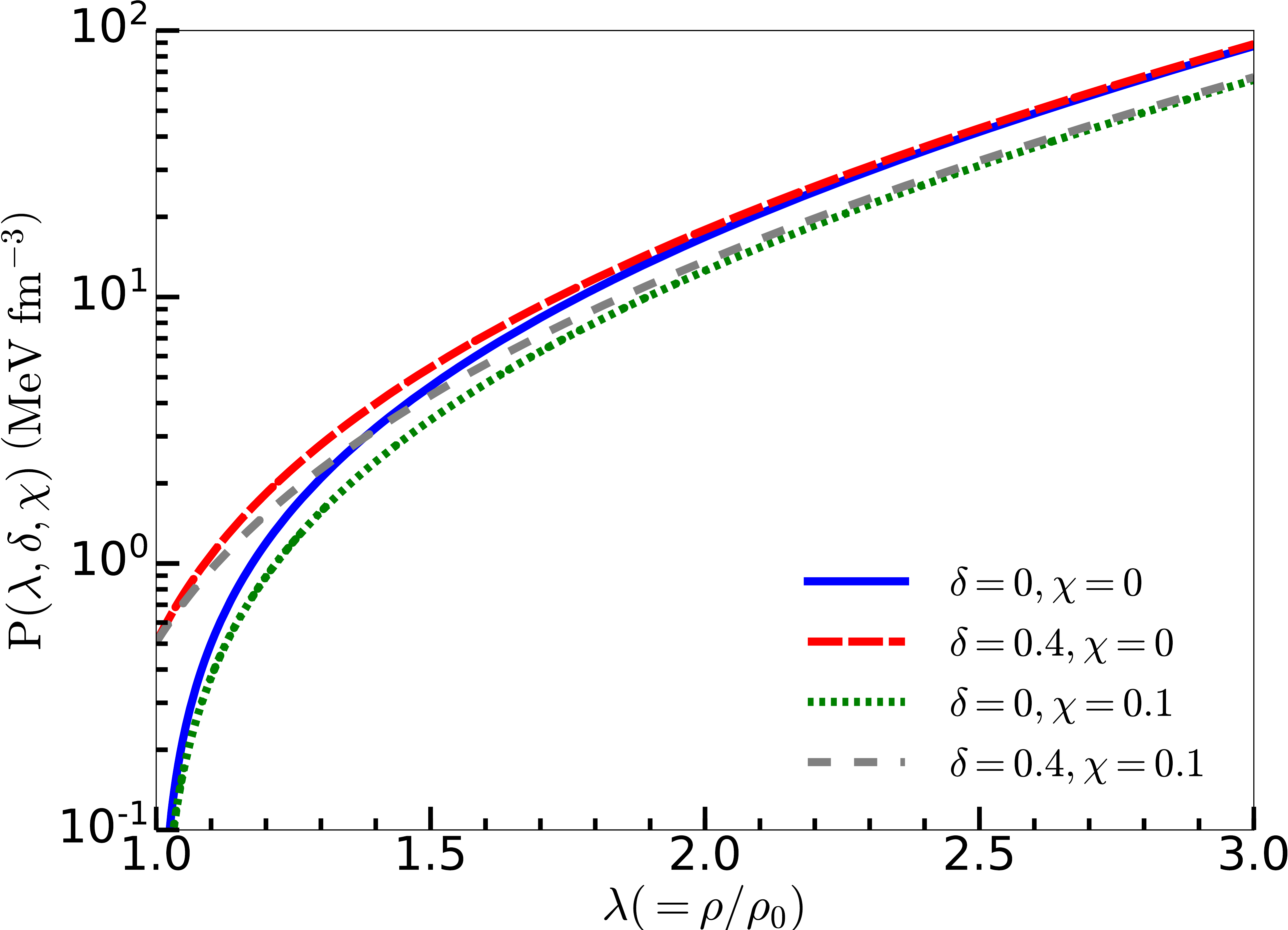}
\caption{Pressure $P(\lambda,\delta,\chi)$ as a function of the
  normalized nuclear matter density $\lambda=\rho/\rho_{0}$.  
  Notations and parameters are the same as in Fig.\,\ref{fig:4}.}
\label{fig:5}
\end{figure}
These results also can be seen from the density dependence of the
pressure shown in Fig.~\ref{fig:5}, where we draw the results for the 
dependence of the pressure on $\lambda$ for possible
four different cases, as discussed in the case of binding energy
dependence on normalized nuclear matter density. One can see that
the strangeness mixing will bring about the softening of EoS, 
comparing the blue solid curve with the green dotted ones or the red
dashed and grey dashed ones.  

\section{Baryons masses in different baryon environments}
\label{sec:4}
We are now in a position to discuss how the masses of the SU(3)
baryons undergo the changes in ordinary and strangeness-mixed nuclear
matter. Since all the medium functions have been already fixed, 
we can study the modification of the baryon masses in different 
nuclear media. While we have considered only the baryon octet in
formulating the nuclear matter, we will investigate the medium
modifications of both the baryon octet and decuplet in nuclear
matter. 

The contributions from the surrounding baryon environment can be
divided into two parts: the change of the classical soliton mass
$\Delta M_{\rm cl}=M_{\rm cl}^*-M_{\rm cl}$ (see
Eq.\,(\ref{eq:collH})) and that of quantum fluctuations 
$\Delta M_{\rm qf}=\Delta M_B-\Delta M_{\rm cl}$, where 
$\Delta M_B=M_B^*-M_B$ denotes the shift in the baryon mass. 
From the values of the parameters in Eq.\,(\ref{eq:CclC1C2}), one can
see that the classical soliton mass is homogeneously dropped in  
nuclear matter. The soliton mass in free space in the present work
is $737.6$~MeV. Its change in the medium at the normal nuclear matter 
density ($\lambda=1$) is given as $-41.38$~MeV. In the case of quantum 
fluctuations the situations is not at all trivial, because different
parts of the quantum fluctuations may behave in a different way
depending on the content of the surrounding baryon environment.  

\begin{figure}[htp]
\centering \includegraphics[scale=0.2]{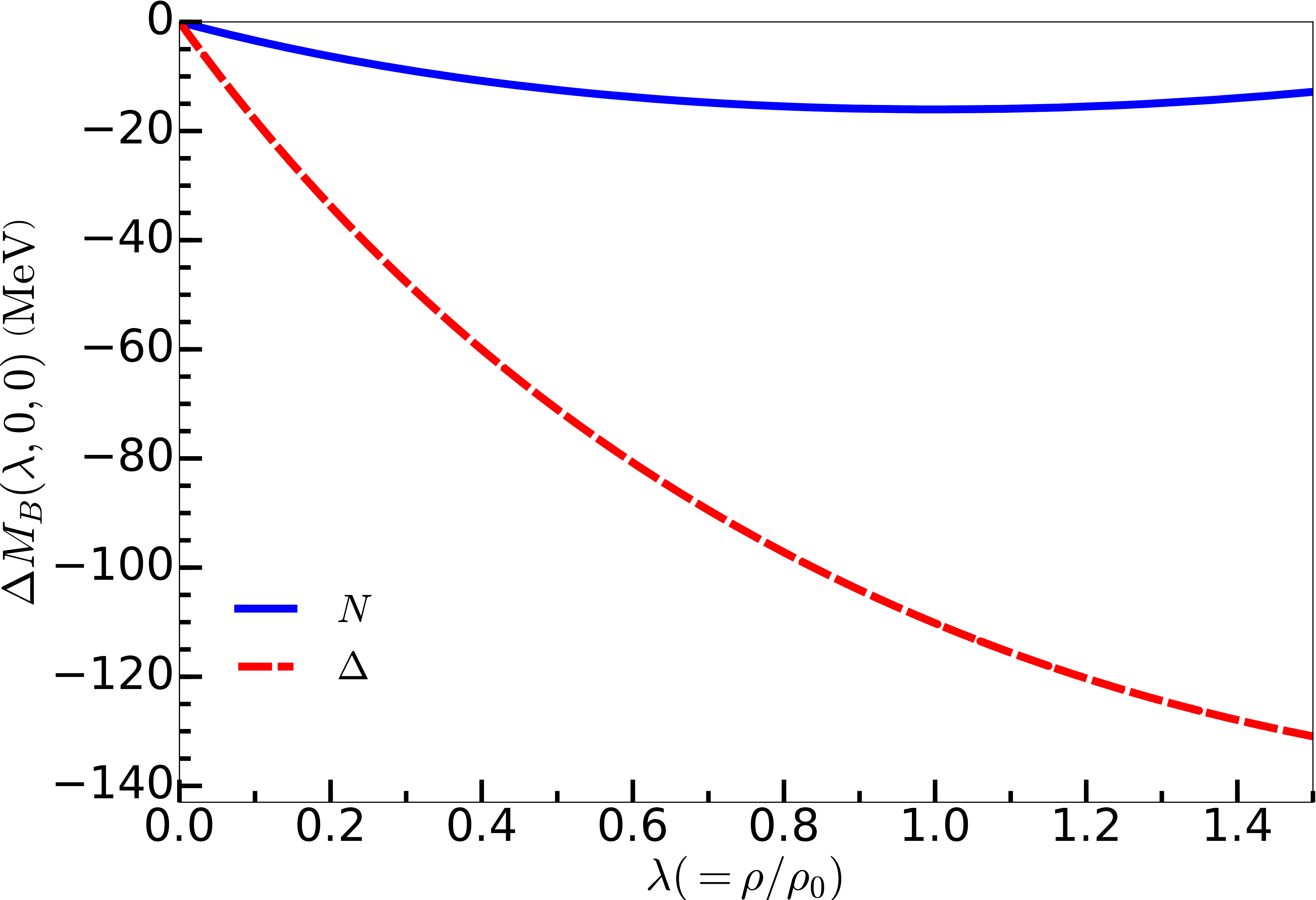}
\caption{The result for the mass shift $\Delta M_B=M^*_B-M_B$ of the
  nucleon $B=N$ in the isospin symmetric nuclear matter is drawn in a
  solid curve, whereas that of the $B=\Delta$ isobar is depicted in a
  dashed one.} 
\label{fig:6}
\end{figure}

The masses of octet and decuplet member baryons in the different 
baryon environments at normal nuclear matter $\lambda=1$ density are
predicted and are listed in Table\,\ref{tab:2}.  
\begin{table*} [hbt]
\caption{Masses of the baryon octet and decuplet both in free space
  and in the different baryon environments at normal nuclear matter 
density $\lambda=1$. The parameters for the symmetry energy
are taken from Set\,I in Table\,\ref{tab:1}. All the masses are given
in units of MeV.} 
\label{tab:2}
\begin{center}
\begin{tabular}{c|cccccc}
\hline \hline
 Baryon  & Exp & Free space &  $\delta=0$, $\chi=0$
 &  $\delta=1$, $\chi=0$ &  $\delta=0$, $\chi=0.1$ & 
     $\delta=0.4$, $\chi=0.1$ \\
   \hline 
  $ p $  & 938.76 & 938.01 &  921.97 & 889.97 & 918.23 & 905.43 \\
  $ n $  & 940.27 & 939.52 & 922.47 & 955.47 & 919.73 & 932.53 \\
  $ \Lambda $ & 1109.61 & 1108.86 &1092.82 & 1092.82 & 1092.77 & 1092.77 \\
 $ \Sigma^{+} $ & 1188.75 & 1188.00 & 1171.96 & 1106.74& 1172.01 & 1145.92 \\
  $ \Sigma^{0} $ & 1190.20 &     1189.45       &     1173.41      &     1173.41       &     1173.46       &     1173.46     \\
 $ \Sigma^{-} $   &     1195.48       &     1194.73       &     1178.69    &     1243.91       &     1178.74       &     1204.83     \\
  $ \Xi^{0} $   &      1319.30       &     1318.55       &     1302.51   &     1269.29       &     1306.21       &     1292.92     \\
  $ \Xi^{-} $  &1321.31 & 1323.78 & 1307.74 & 1340.96 & 1311.40 &1324.72 \\
  $ \Delta^{++} $ \ &   1230.55  &   1247.79  &   1137.60 &   1041.30  &   1133.94  &   1095.42\\
   $ \Delta^{+} $ &   1234.90  &   1248.61  & 1138.42
  &   1106.31    &   1134.75    & 1121.91  \\
  $ \Delta^{0} $ &   1231.3&   1250.79&   1140.59
  &   1172.70&   1136.93  &  1149.77  \\
  $ \Delta^{-} $  &   1230 to 1234 &   1254.33  &   1144.14
 & 1240.44 &   1140.48 &   1179.00   \\
   $ \Sigma^{*+} $ & 1382.80  &   1387.73  &  1277.54    & 1213.34 &   1277.54  &   1251.86  \\
  $ \Sigma^{*0} $ &   1383.70  &   1389.91 &   1279.72
  &   1279.72&   1279.72   &   1279.72  \\
  $ \Sigma^{*-} $ \ &   1394.20  &   1393.46 &   1283.27
   &   1374.47 &   1283.27  &   1308.95 \\
  $ \Xi^{*0} $ & 1531.80 &   1529.03 &   1418.84 &1386.74  &   1422.50  &   1409.66   \\
  $ \Xi^{*-} $  & 1535.0  &   1532.58 &   1422.39
  & 1454.49    &   1426.05   &   1438.89   \\
$ \Omega^{-} $  & 1672.45   &   1671.70  &   1561.51  & 1561.51    &   1568.84    &   1568.84  \\
\hline \hline
\end{tabular}
\end{center}
\end{table*}
One can see that the change of quantum fluctuations and,
consequently, the changes of baryons masses in the 
different environments are different. In the symmetric 
ordinary nuclear matter ($\lambda=1$ and $\delta=0$)
the masses of the baryon octet and decuplet decrease as $\lambda$
increases (compare third and fourth columns in Table~\ref{tab:2}).
Figure~\ref{fig:6} illustrates the density 
dependence of the mass shifts of the nucleon and $\Delta$ isobar in 
the isospin symmetric nuclear matter. 
As shown in Fig.~\ref{fig:6}, the mass shift of the nucleon decreases
very slowly as $\lambda$ increases. However, it is almost saturated in
the vicinity of the normal nuclear matter density and then starts to
increase very slightly. On the other hand, the mass shift of the
$\Delta$ isobar falls off monotonically as $\lambda$ increases. 
Since the mass difference of the nucleon and $\Delta$ comes from the
zero-mode quantization of the chiral soliton, the medium modification
of the zero-mode quantum fluctuation for the $\Delta$ comes into
essential play. This fact makes the mass shift of $\Delta$ turn out
to be very different from that of the nucleon. We find the very
similar results for the other members of the baryon octet and
decuplet. 

However, the situation is changed in isospin-asymmetric matter. The
mass shift of the SU(3) baryons in the isospin asymmetric environment
depends on the third component of baryons isospin. Thus, the mass
shifts of the baryons are more pronounced, in particular, for the
baryon with negative $T_3$. For example, 
the mass shift of the proton in pure neutron matter ($\delta=1$) at
normal nuclear matter density ($\lambda=1$) is obtained to be
$-48.79$\,MeV, while that of the neutron becomes positive,
i.e. $+15.20$\,MeV (see also Table~\ref{tab:2}). This implies that the
up and down quarks may undergo changes in a different manner.  

\begin{figure}[htp]
\centering 
\includegraphics[scale=0.2]{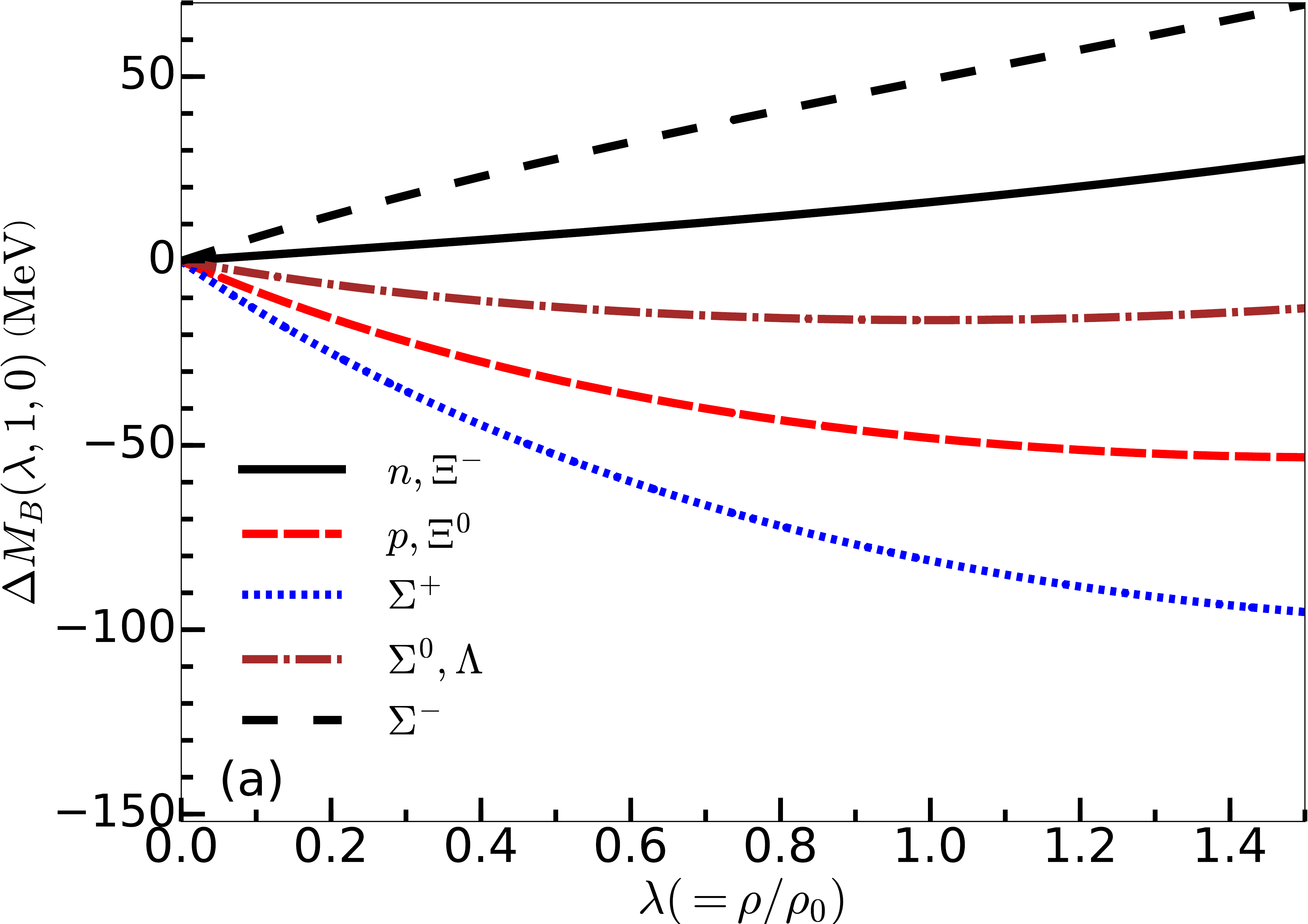}\hspace{0.5cm}
\includegraphics[scale=0.2]{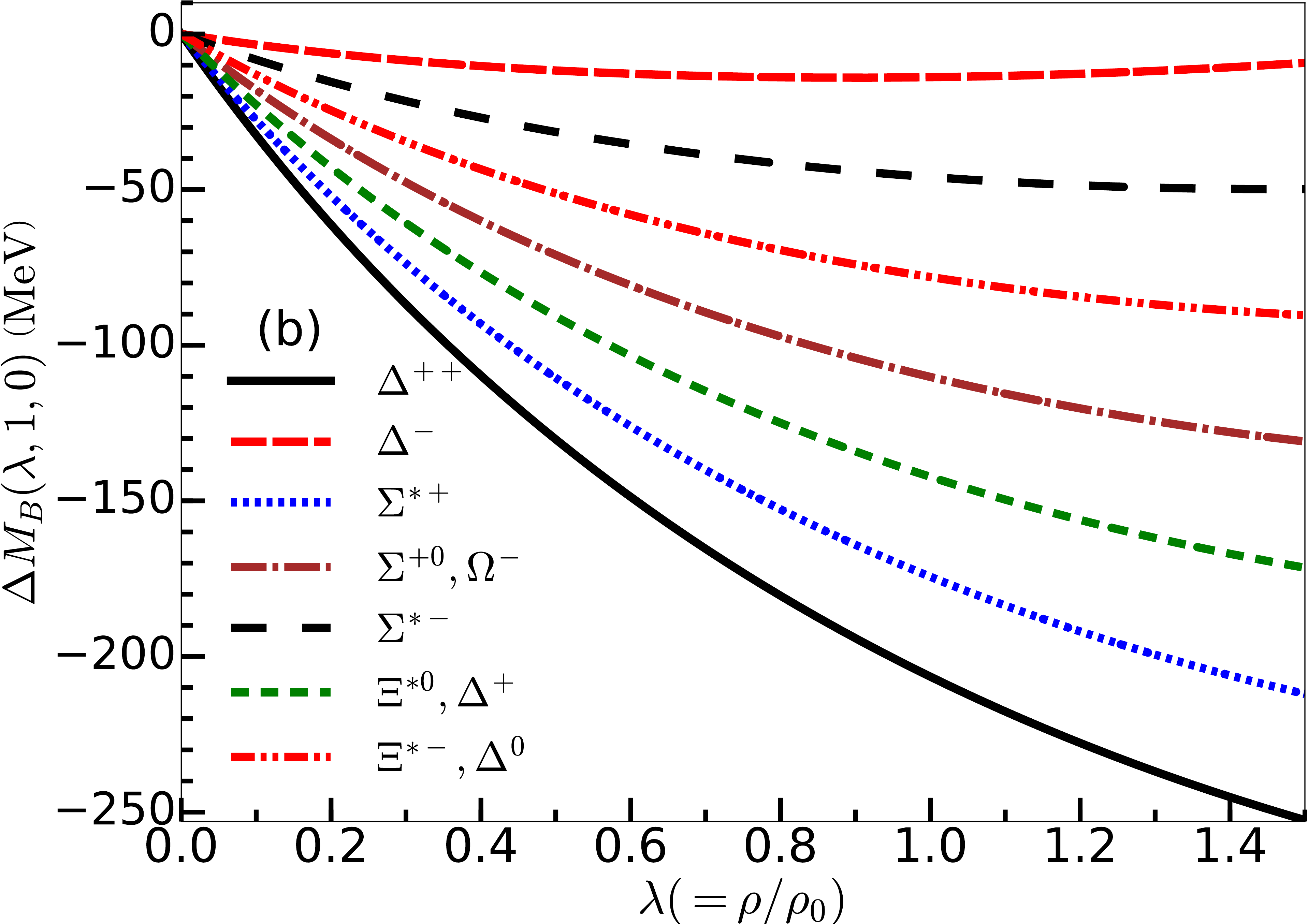}
\caption{Results for the mass shifts $\Delta M_B$ of the baryon octet
  and the decuplet in the pure neutron matter ($\delta=1$) are drawn
  in the upper (a) and lower panels (b), respectively. The parameters
  for the symmetry energy are taken from Set\,I in Table\,\ref{tab:1}.}
\label{fig:7}
\end{figure}
The results for the mass shifts of the baryon octet and decuplet in
the pure neutron matter are shown in the upper and lower panels of
Fig.~\ref{fig:7}, respectively. First of all, one can explicitly see
that the effects of the isospin mass splitting are clearly shown 
in the isospin-asymmetric nuclear environment.
Depending on the charges of the baryon octet, we can see that their
mass shifts behave in a very different way. As shown in the upper
panel of Fig.~\ref{fig:7}, the neutron mass increases as $\lambda$
increases, whereas the proton mass drops off as $\lambda$ increases. 
In general, the masses of the members in the baryon octet 
with the negative values of $T_3$ rises in the neutron matter as
$\lambda$ increases. 
On the other hand, the masses of the octet baryons with the positive
values of $T_3$ drop off as $\lambda$ decreases. However, while the
mass of $\Xi^0$ is identical to that of the proton, those of $\Sigma^0$
and $\Lambda^0$, which are also identical each other, fall off slowly
and then are saturated as $\lambda$ increases. 
This is originated from the fact that the in-medium functionals for 
the quantum fluctuations near the third component of isospin are quite
sensitive to the medium effects and they are identical for  
the baryons that have the same isospin components.

The general tendency and the effects of the isopin factor can be seen 
also in the case of decuplet baryons, which are given in the lower (b)
panel of Fig.~\ref{fig:7}, but the mass shifts are larger than 
those of the baryon octet.  For example, the mass shift in the isospin
averaged $\Delta$ in normal nuclear matter is around two times 
smaller than the $\Delta^{++}$ mass shift in neutron matter at normal
nuclear matter density $\lambda=1$. For example, one can see this by
comparing the red dashed curve in Fig.~\ref{fig:6} with the black
solid one in the lower panel of Fig.~\ref{fig:7}. The isospin
component factor is similar to the octet case and some baryons masses
such as $\Xi^{*0}$ and $\Delta^+$ have the identical dependence on
$\lambda$. Due to the isospin factor, the $\Delta^{-}$ mass in neutron 
matter remains almost constant which is seen from the red dashed curve
in the lower panel of Fig.~\ref{fig:7}. The present results are
in qualitative agreement with those from Ref.~\cite{Savage:1995kv}. 

\begin{figure}[htp]
\centering 
\includegraphics[scale=0.2]{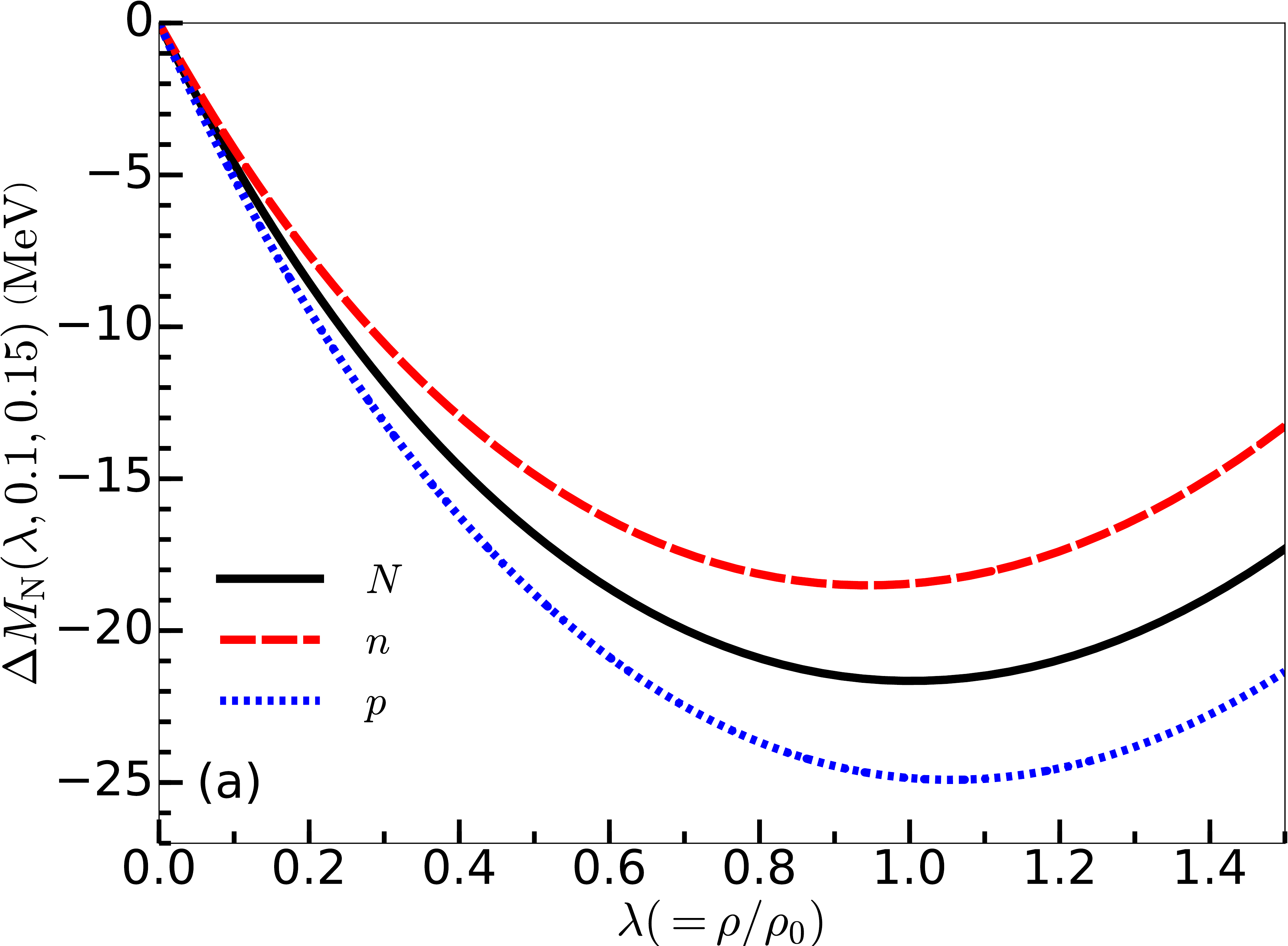}\hspace{0.5cm}
\includegraphics[scale=0.2]{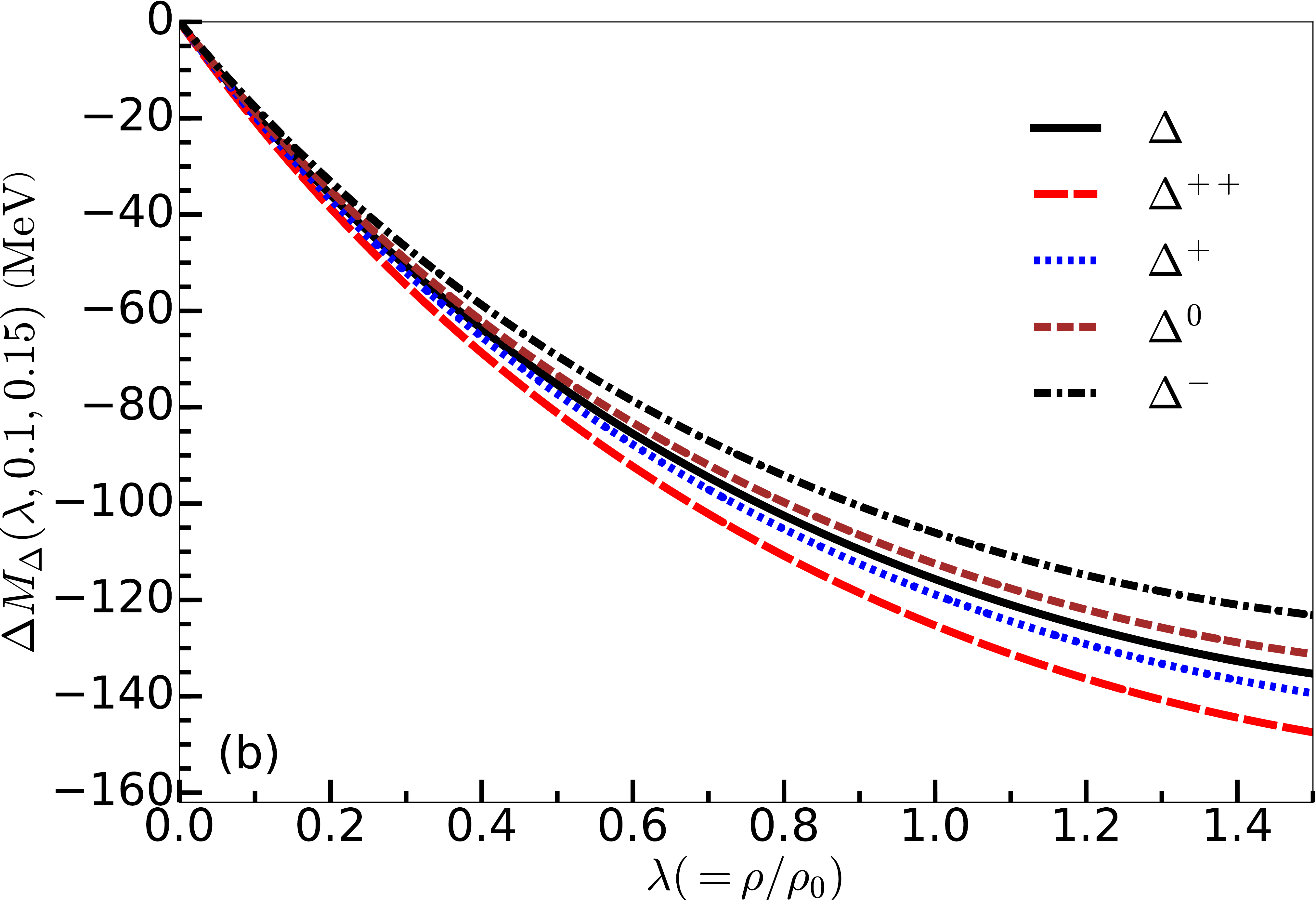}
\caption{Mass shifts $M^*_B-M_B$ of the nucleon
(upper (a) panel)  and $\Delta$ 
isobar (lower (b) panel) in
 the strangeness mixed isospin-asymmetric matter. } 
\label{fig:8}
\end{figure}
For completeness, we show also the mass changes of the baryons in the
strangeness-mixed asymmetric environment. In Fig.~\ref{fig:8}, the
results for the mass shifts of the nucleon and $\Delta$ in
strangeness-mixed asymmetric matter are presented as functions of
$\lambda$, which we choose them as the representatives of the baryon
octet and decuplet, respectively. For the sake of the illustration, we
include the changes of isospin-averaged masses of the nucleon and $\Delta$
by the black solid curves in the upper and lower panels,
respectively. Figure~\ref{fig:8} explicitly shows the isospin
factor in nuclear matter, which are explained above.
It also depicts how the strength of the mass are changed due to the
environment content. 
Comparing the Fig.~\ref{fig:7} with Fig.~\ref{fig:8},
one can conclude that the changes in neutron matter 
are stronger than those in strange matter.
In general, the results in strangeness-mixed matter are
rather similar to those in pure neutron matter.

\section{Summary and outlook}
\label{sec:5}

In the present work, we have investigated the various baryonic matters
such as the symmetric nuclear matter, pure neutron matter, and 
strangeness-mixed baryonic matter, based on the meson mean-field
approach or the generalized SU(3) chiral soliton model. 
All the dynamical parameters for the baryon masses in the model
were determined by using the experimental data in free space and then 
we have introduced the parametrizations for the density-dependent
parameters. As a starting point, we took a "\emph{model-independent
  approach}" for the SU(3) baryon properties in free space, which
described successfully the baryon masses of the baryon decuplet and other
properties of baryons in free space~\cite{Yang:2010fm,
  Yang:2015era}. In the present work, the medium modifications of the
model functionals were carried out by employing the linear
density-dependent forms. Having determined 
the parameters for the medium modification by using the empirical data 
related to nuclear matter such as the binding energy per nucleon,
the compressibility, and the symmetry energy, we were able to describe
the equation of states for various nuclear environments 
including the nonstrange sector. We found that the present results
were in good 
agreement with the data extracted from the phenomenology and
experiments and with the results from other approaches. We also
discussed the properties of the strangeness-mixed matter and they also
were in agreement with the phenomenology. 
Finally, we predicted the mass shifts of the baryon octet and 
decuplet in various baryonic environments with different content of
the isospin asymmetry and strangeness. We scrutinized the changes of
the masses of the baryons with different values of the third components
of isospin and found that the masses of the baryons with negative
charges show very different dependence on the nuclear matter density
from those with positive and null charges. 

Since we have formulated the equations of states for
isospin-asymmetric and strangeness-mixing baryonic matter,
one can directly apply the present model to investigate 
properties of neutron stars. The corresponding investigation is under
way. 

\section*{Acknowledgments}
The present work was supported by Basic Science Research Program through the 
National Research Foundation (NRF) of Korea funded by the Korean government 
(Ministry of Education, Science and Technology, MEST), Grant Numbers 
2019R1A2C1010443 (Gh.-S. Y.), 2018R1A2B2001752 and 2018R1A5A1025563 
(H.-Ch.K.),  2020R1F1A1067876 (U. Y.).

\appendix

\section{Masses of baryons in free space}
\label{app:A}

The masses of baryon octet are expressed as 
\begin{align}
M_{N} & =  M_{{\rm cl}}+E_{(1,\,1),\,1/2}
+\frac{1}{5}\left(c_{8}+\frac{4}{9}c_{27}\right)T_{3}\cr
&+\frac{3}{5}\left(c_{8}+\frac{2}{27}c_{27}\right)
      \left(T_{3}^{2}+\frac{1}{4}\right)\cr
 &-\left(d_{1}-d_{2}\right)T_{3}-\left(D_{1}+D_{2}\right),
\\
M_{\Lambda} 
& =  M_{{\rm cl}}+E_{(1,\,1),\,1/2}
+\frac{1}{10}\left(c_{8}-\frac{2}{3}c_{27}\right)\cr
&-D_{2},
\\
M_{\Sigma} 
& =  M_{{\rm cl}}+E_{(1,\,1),\,1/2}
+\frac{1}{2}\left(T_{3}-\frac{1}{5}\right)c_{8}\cr
&+\frac{2}{9}\left(T_{3}^{2}-\frac{7}{10}\right)c_{27}
-\left(d_{1}+\frac{1}{2}d_{2}\right)T_{3}\cr
&+D_{2},\\
M_{\Xi} 
& =  M_{{\rm cl}}+E_{(1,\,1),\,1/2}
+\frac{4}{5}\left(c_{8}-\frac{1}{9}c_{27}\right)T_{3}\cr
&-\frac{2}{5}\left(c_{8}-\frac{1}{9}c_{27}\right)
\left(T_{3}^{2}+\frac{1}{4}\right)\cr
&
-\left(d_{1}+2d_{2}\right)T_{3}\;+\;D_{1},
\end{align}
where $E_{(1,1),1/2}$ can be obtained from Eq.~\eqref{eq:EpqJ}. 
The masses of the baryon decuplet are given by the following
expressions 
\begin{align}
M_{\Delta} 
& =  
M_{{\rm cl}}+E_{(3,\,0)\,3/2}
+\frac{1}{4}\left(c_{8}+\frac{8}{63}c_{27}\right)T_{3}\cr
&+\frac{5}{63}T^{2}_{3}
+\frac{1}{8}\left(c_{8}-\frac{2}{3}c_{27}\right)\cr
&
-\left(d_{1}-\frac{3}{4}d_{2}\right)T_{3} 
- \left(D_{1}-\frac{3}{4}D_{2}\right), \\
M_{\Sigma^{\ast}} 
& = M_{{\rm cl}}+E_{(3,\,0),\,3/2}
+\frac{1}{4}\left(c_{8}-\frac{4}{21}c_{27}\right)T_{3}\cr
&+\frac{5}{63}c_{27}\left(T_{3}^{2}-1\right)
-\left(d_{1}-\frac{3}{4}d_{2}\right)T_{3},\\
M_{\Xi^{\ast}} 
& = 
M_{{\rm cl}}+E_{(3,\,0),\,3/2}
+\frac{1}{4}\left(c_{8}-\frac{32}{63}c_{27}\right)T_{3}\cr
&-\frac{1}{4}\left(c_{8}+\frac{8}{63}c_{27}\right) 
\left(T_{3}^{2}+\frac{1}{4}\right)\cr
&
-\left(d_{1}-\frac{3}{4}d_{2}\right)T_{3}
+\left(D_{1}-\frac{3}{4}D_{2} \right),\\
M_{\Omega} 
& =M_{{\rm cl}}+E_{(3,\,0),\,3/2}
-\frac{1}{4}\left(c_{8}-\frac{4}{21}c_{27}\right)\cr
&+2\left(D_{1}-\frac{3}{4}D_{2} \right),
\end{align}
where $E_{(3,0),3/2}$ can be obtained from Eq.~\eqref{eq:EpqJ}.
Here $d_{1,2}$ and $D_{1,2}$ are defined as 
\begin{align}
\label{eq:d1}
d_{1} & =
        (m_{d}-m_{u})\left[-\frac{1}{5}\alpha-\beta+\frac{1}{5}\gamma
        \right],\\ 
d_{2} & =  (m_{d}-m_{u})\left[-\frac{1}{10}\alpha-\frac{3}{20}
\gamma\right],\\
D_{1} & =
        (m_{s}-\bar{m})\left[-\frac{1}{5}\alpha-\beta+\frac{1}{5}\gamma
        \right],\\ 
D_{2} & =
        (m_{s}-\bar{m})\left[-\frac{1}{10}\alpha-\frac{3}{20}\gamma\right]. 
\label{eq:cD2}
\end{align}
The explicit forms of $c_8$ and $c_{27}$, which denote the
wave-function corrections, can be found in 
Ref.\,\cite{Yang:2010fm}.

\end{document}